\documentclass[11pt,onecolumn,draft]{IEEEtran} 

\usepackage[final]{graphicx}
\usepackage{graphicx}
\usepackage{amsfonts,amssymb}
\usepackage{latexsym}
\usepackage{amsmath}   

\newtheorem{thm}{Theorem}[section]

\newtheorem{lem}[thm]{Lemma}

\newtheorem{assumption}[thm]{Assumption}

\newtheorem{definition}[thm]{Definition}

\newtheorem{example}[thm]{Example}

\newtheorem{remark}[thm]{Remark}

\newcommand{\bi}{\begin{itemize}}
\newcommand{\ei}{\end{itemize}}
\newcommand{\ben}{\begin{enumerate}}
\newcommand{\een}{\end{enumerate}}
\newcommand{\beq}{\begin{equation}}
\newcommand{\eeq}{\end{equation}}
\newcommand{\beqa}{\begin{eqnarray}}
\newcommand{\eeqa}{\end{eqnarray}}

\def\BibTeX{{\rm B\kern-.05em{\sc i\kern-.025em b}\kern-.08em
    T\kern-.1667em\lower.7ex\hbox{E}\kern-.125emX}}

\begin{document}

\title{Tree-Based Construction of LDPC Codes Having Good Pseudocodeword 
Weights}

\author{\begin{tabular}{cc}
\vspace{-0.00in}Christine A. Kelley&\vspace{-0.00in}Deepak Sridhara and Joachim Rosenthal\\
\vspace{-0.00in}Department of Mathematics& \vspace{-0.00in} Institut f$\ddot{\mbox{u}}$r Mathematik\\ 
\vspace{-0.00in}University of Notre Dame& \vspace{-0.00in}Universit$\ddot{\mbox{a}}$t Z$\ddot{\mbox{u}}$rich \\
\vspace{-0.00in}Notre Dame, IN 46556, U.S.A.& \vspace{-0.00in}CH-8057, Z$\ddot{\mbox{u}}$rich, Switzerland.\\
\vspace{-0.00in}email: {\tt ckelley1}@nd.edu& \vspace{-0.00in}email: \{{\tt
sridhara, rosen}\}@math.unizh.ch
\end{tabular}\thanks{This work was supported in part by the NSF Grant No. CCR-ITR-02-05310. Some of the material in this paper was previously presented at ISIT 2005 (Adelaide, Australia).}
\centerline{(Submitted Oct. 3, 2005; Revised May 1, 2006, Nov. 25, 2006.)}
}

\markboth{Submitted to IEEE Transactions on Information
  Theory}{Kelley et. al}

\maketitle

\vspace{-0.51in}
\begin{abstract}
We present a tree-based construction of LDPC codes that have minimum
pseudocodeword weight equal to or almost equal to the minimum distance, and perform
well with iterative decoding. The construction involves
enumerating a $d$-regular tree for a fixed number of layers and
employing a connection algorithm based on permutations or
mutually orthogonal Latin squares to close the tree. Methods are
presented for degrees $d=p^s$ and $d = p^s+1$, for $p$ a
prime. One class corresponds to the well-known finite-geometry
and finite generalized quadrangle LDPC codes; the other codes presented
are new.  We also present some bounds on
pseudocodeword weight for $p$-ary LDPC codes. Treating these
codes as $p$-ary LDPC codes rather than binary LDPC codes
improves their rates, minimum distances, and pseudocodeword weights, thereby giving a new
importance to the finite geometry LDPC codes where $p > 2$.
\end{abstract}


\section{Introduction}
Low Density Parity Check (LDPC) codes are widely acknowledged to be good 
codes due to their near Shannon-limit performance when decoded 
iteratively. However, many structure-based constructions of LDPC codes 
fail to achieve this level of performance, and are often outperformed by 
random constructions.  (Exceptions include the finite-geometry-based LDPC 
codes (FG-LDPC) of \cite{ko01}, which were later generalized in \cite{li05}.) 
Moreover, there are discrepancies between iterative and maximum likelihood 
(ML) decoding performance of short to moderate blocklength LDPC codes. 
This behavior has recently been attributed to the presence of so-called 
{\em pseudocodewords} of the LDPC constraint graphs (or, Tanner graphs), which are valid 
solutions of the iterative decoder which may or may not be optimal 
\cite{ko03p}. Analogous to the role of minimum Hamming distance, 
$d_{\min}$, in ML-decoding, the minimum pseudocodeword weight\footnote{Note that the minimum
pseudocodeword weight is specific to the LDPC graph representation of the LDPC code.}, $w_{\min}$, 
has been shown to be a leading predictor of performance in iterative 
decoding \cite{wi96t}. Furthermore, the error floor performance of iterative decoding 
is dominated by minimum weight pseudocodewords. Although there exist 
pseudocodewords with weight larger than $d_{\min}$ that have adverse 
affects on decoding, it has been observed that
pseudocodewords with weight $w_{\min} < d_{\min}$ are 
especially problematic \cite{ke05u}.

Most methods for designing LDPC codes are based on random
design techniques. However,
the lack of structure implied by this randomness
presents serious disadvantages in terms of
storing and accessing a large parity check
matrix, encoding data, and analyzing code performance. Therefore,
by designing codes algebraically, some of these problems
can be overcome. In the recent literature, several algebraic
methods for constructing LDPC codes have been proposed
\cite{ta04}\cite{ko01}\cite{ro00p}\cite{ka03}. These constructions
are geared towards optimizing a specific parameter in the design
of Tanner graphs -- namely, either 
girth, expansion,  diameter, or more recently, stopping
sets. In this paper, we consider a more fundamental parameter for
designing LDPC codes -- namely, pseudocodewords of the
corresponding Tanner graphs. While pseudocodewords are
essentially stopping sets on the binary erasure channel (BEC) and have
been well studied on the BEC in \cite{di02, ka03, sc06,ha05u}, they
have received little attention in the context of designing LDPC
codes for other channels. The constructions presented in this
paper are geared towards maximizing the minimum pseudocodeword
weight of the corresponding LDPC Tanner graphs.

The Type I-A construction and certain cases of the Type II construction 
presented in this paper are designed so that the resulting codes have 
minimum pseudocodeword weight equal to or almost equal to the minimum distance of the code, and consequently, the problematic low-weight pseudocodewords are 
avoided.  Some of the resulting codes have minimum distance which meets
 the lower tree bound originally presented in \cite{ta81a}. Since 
$w_{\min}$ shares the same lower bound \cite{ke05u,ke04p}, and is upper 
bounded by $d_{\min}$, these constructions have $w_{\min} = d_{\min}$. It 
is worth noting that this property is also a characteristic of some of the 
FG -LDPC codes \cite{li05}, and indeed, the projective-geometry-based 
codes of \cite{ko01} arise as special cases of our Type II construction. 
It is worth noting, however, that the tree construction technique is 
simpler than that described in \cite{ko01}. Furthermore, the Type I-B 
construction presented herein
 yields a family of codes with a wide 
range of rates and blocklengths that are comparable to those obtained from 
finite geometries. This new family of codes has $w_{\min} =
d_{\min} \ge $ tree bound in most cases. 

Both min-sum and sum-product iterative decoding performance of the tree-based
constructions are comparable to, if not, better than, that of random LDPC
codes of similar rates and block lengths. We now present the tree
bound on $w_{\min}$  derived in \cite{ke04p}. 

\begin{definition}
The tree bound of a $d$ left (variable node) regular bipartite
LDPC constraint graph with girth $g$ is defined as 
\begin{equation}
T(d,g):= \left\{\begin{array}{cc}
1+d + d(d-1) + d(d-1)^2+\ldots+ d(d-1)^{\frac{g-6}{4}}, & \frac{g}{2}\mbox{ odd },\\
1+ d+d(d-1)+\ldots+d(d-1)^{\frac{g-8}{4}} + (d-1)^{\frac{g-4}{4}},& \frac{g}{2}\mbox{ even }.
\end{array}\right . 
\end{equation}
\label{treebound_defn}
\end{definition}

\begin{thm}
{ \em Let $G$ be a bipartite LDPC constraint graph with smallest left (variable node) degree $d$ and girth $g$. Then the minimum pseudocodeword weight $w_{\min}$ (for the AWGN/BSC channels) is lower bounded by
\vspace{-0in}
 \[\scriptsize w_{\min} \ge T(d,g). \]
}
\label{thm1}
\end{thm}

This bound is also the tree bound  on the minimum distance
established by Tanner in \cite{ta81a}. And since the set of
pseudocodewords includes all codewords, we have $w_{\min}\le
d_{\min}$. 

 We derive a pseudocodeword weight definition for the $p$-ary symmetric channel (PSC), and extend
the tree lower bound on $w_{\min}$ for the PSC. The tree-based code constructions are
then analyzed as $p$-ary LDPC codes. Interpreting the tree-based codes as $p$-ary LDPC codes when
the degree is $d = p^s$ or $d = p^s+1$ yields codes with rates $> 0.5$ and good distances. The
interpretation is also meaningful for the FG LDPC codes of \cite{ko01}, 
since the projective
geometry codes with $d = p^s+1, p>2$ have rate $\frac{1}{n}$ if treated as binary codes and rate $>
0.5$ if treated as $p$-ary LDPC codes.

The paper is organized as follows.  The following section introduces permutations
and mutually orthogonal Latin squares. The Type I constructions are presented in Section 3
and properties of the resulting codes are discussed. Section 4
presents the Type II construction with two variations and the
resulting codes are compared with codes arising from finite
geometries and finite generalized quadrangles.  In Section 5, we provide simulation
results of the codes on the additive white Gaussian noise (AWGN)
channel and on the p-ary symmetric channel. The paper is
concluded in Section 6. \\ \vspace{0.1in}

\section{Preliminaries}

\subsection{Permutations}

A permutation on set of integers modulo $m$, $\{0,1,\dots,m-1\}$ is 
a bijective map of the form 
\[\pi:\{0,1,\dots,m-1\}\rightarrow \{0,1,\dots,m-1\}\]
A permutation is commonly denoted either as 
\[\left( \begin{array}{ccccc}
0 & 1 & 2&\dots& m-1\\
\pi(0)&\pi(1)&\pi(2)&\dots&\pi(m-1)
\end{array}\right) \]
or as $(a_{11} a_{12} \dots a_{1s_1}) (a_{21} a_{22} \dots 
a_{2s_2})\dots$,
where $a_{i2}=\pi(a_{i1}), a_{i3}=\pi(a_{i2}),\dots, 
a_{is_i}=\pi(a_{i,s_i-1}), a_{i1}=\pi(a_{is_i})$ for all $i$.

As an example, suppose $\pi$ is a permutation on the set $\{0,1,2,3\}$ given 
by $\pi(0)=0$, $\pi(1)=2$, $\pi(2)=3$, $\pi(3)=1$. Then $\pi$ is denoted
as $\left(\begin{array}{cccc}
0&1&2&3\\
0&2&3&1
\end{array}\right)$ in the former representation, and as 
$(0)(123)$ in the latter representation.

\subsection{Mutually orthogonal Latin squares (MOLS)}
Let $\mathbb{F}:=GF(q)$ be a finite field of order $q$ and let
$\mathbb{F}^{*}$ denote the corresponding multiplicative group
--i.e., $\mathbb{F}^{*}:=\mathbb{F}\backslash \{0\}$. For every
$a\in \mathbb{F}^{*}$, we define a $q\times q$ array having
entries in $\mathbb{F}$ by the following linear map\[
\phi_a: \mathbb{F}\times \mathbb{F}\rightarrow
\mathbb{F}\]
\[(x,y)\mapsto x+a\cdot y \]
where `$+$' and `$\cdot$' are the corresponding field
operations. The above set of maps define $q-1$ mutually
orthogonal Latin squares (MOLS) [pg. 182 -- 199, \cite{li01b}]. 
The map $\phi_a$ can be written as a matrix $M_a$ where the rows and columns of
the matrix are indexed by the elements of $\mathbb{F}$ and the $(x,y)^{th}$
entry of the matrix is $\phi_a(x,y)$.
By introducing another map $\phi_0$ defined in the following manner
\[ \phi_0: \mathbb{F}\times \mathbb{F}\rightarrow \mathbb{F}\]
\[(x,y)\mapsto x \]
we obtain an additional array $M_0$ which is orthogonal to the above
family of $q-1$ MOLS. However, note that $M_0$ is not a Latin
square. We use this set of $q$ arrays in the subsequent
tree-based constructions. 

As an example, let $\mathbb{F} =\{0,1,\alpha,\alpha^2\}$ be the finite field
with four elements, where $\alpha$ represents the primitive element. 
Then, from the above set of maps we obtain the following four
orthogonal squares

{\tiny
\[
M_{0}:=\hspace{-0in} \left[\begin{array}{cccc}
0&0&0&0\\
1&1&1&1\\
\alpha&\alpha&\alpha&\alpha\\
\alpha^2&\alpha^2&\alpha^2&\alpha^2
\end{array}\right], \ M_{1}:=\left[\begin{array}{cccc}
0&1&\alpha&\alpha^2\\
1&0&\alpha^2&\alpha\\
\alpha&\alpha^2&0&1\\
\alpha^2&\alpha&1&0
\end{array}\right], \ M_{\alpha}:=\left[\begin{array}{cccc}
0&\alpha&\alpha^2&1\\
1&\alpha^2&\alpha&0\\
\alpha&0&1&\alpha^2\\
\alpha^2&1&0&\alpha
\end{array}\right], \ M_{\alpha^2}:=\left[\begin{array}{cccc}
0&\alpha^2&1&\alpha\\
1&\alpha&0&\alpha^2\\
\alpha&1&\alpha^2&0\\
\alpha^2&0&\alpha&1
\end{array}\right] .\]}

\vspace{0.1in}

\section{Tree-based Construction: Type I}
In the Type I construction, first a $d$-regular tree of 
alternating ``variable'' and ``constraint'' node layers is 
enumerated downwards from a root variable node (layer $ L_0$) 
for $\ell$ layers. (The variable nodes and constraint nodes in this tree 
are merely two different types of vertices that give rise to a  bipartition in the graph.) 
If $\ell$ is odd (respectively, even), the final 
layer $L_{\ell - 1}$ is composed of variable nodes 
(respectively, constraint nodes). Call this tree $T$. The tree $T$ is 
then reflected across an imaginary horizontal axis to yield another tree, 
$T'$,  and  the variable and constraint nodes are reversed. 
That is, if layer $L_i$ in $T$ is composed of variable nodes, 
then the reflection of $L_i$, call it $L_i'$, is 
composed of constraint nodes in the reflected tree, $T'$. 
The union of these two trees, along with edges connecting the nodes in 
layers $L_{\ell-1}$ and $L_{\ell - 1}'$ according to a connection 
algorithm that is described next, comprise the graph representing a 
Type I LDPC code. We now present two connection schemes that can be 
used in this Type I model, and discuss the resulting codes.\\

\subsection{Type I-A}
Figure~\ref{type1Ad3g10_graph} shows a 3-regular girth $10$ Type I-A LDPC
constraint graph. For $d=3$, the Type I-A construction yields a $d$-regular LDPC
constraint graph  having $1+d+d(d-1)+\ldots+d(d-1)^{\ell-2}$
variable and constraint nodes, and girth $g$.  The tree $T$ has
$\ell$ layers. To connect the nodes in $L_{\ell - 1}$ to $L_{\ell
  - 1}'$, first label the variable (resp., constraint) nodes in
$L_{\ell-1}$ (resp., $L_{\ell-1}'$) when $\ell$ is odd (and vice
versa when $\ell$ is even), as $v_0,v_1,\dots,v_{2^{\ell-2}-1}$,
$v_{2^{\ell-2}},\dots,v_{2\cdot 2^{\ell-2}-1},v_{2\cdot
  2^{\ell-2}},\dots,v_{3\cdot 2^{\ell-2}-1}$ (resp.,
$c_0,c_1,\dots,c_{3\cdot 2^{\ell-2}-1}$). The nodes
$v_0,v_1,\dots,v_{2^{\ell-2}-1}$ form the $0^{th}$ class $S_0$,
the nodes $v_{2^{\ell-2}},\dots,v_{2\cdot 2^{\ell-2}-1}$ form the
$1^{st}$ class  $S_1$, and the nodes $v_{2\cdot
  2^{\ell-2}},\dots,v_{3\cdot 2^{\ell-2}-1}$ form the $2^{nd}$
class $S_2$; classify the constraint nodes into $S_0'$, $S_1'$,
and $S_2'$ in a similar manner. In addition, define four permutations $\pi(\cdot), \tau(\cdot),\tau'(\cdot), \tau''(\cdot)$ of the set $\{0,1,\dots,2^{\ell-2}-1\}$ and connect the nodes in
 $L_{\ell - 1}$ to  $L_{\ell - 1}'$  as follows: For $j=0,1,\dots,2^{\ell-2}-1$,
\begin{enumerate}
\item The variable node $v_{j}$ is connected to nodes $c_{\pi(j)}$ and $c_{\tau(j)+ 2^{\ell-2}}$.
\item The variable node $v_{j+ 2^{\ell-2}}$ is connected to nodes $c_{\pi(j)+ 2^{\ell-2}}$ and $c_{\tau'(j)+ 2\cdot 2^{\ell-2}}$.
\item The variable node $v_{j+2\cdot 2^{\ell-2}}$ is connected to nodes $c_{\pi(j)+2\cdot 2^{\ell-2}}$ and $c_{\tau''(j)}$.
\end{enumerate}

The permutations for the cases $g=6,8,10,12$ are given in
Table~\ref{table_typeIapermutations}. For $\ell=3,4,5,6$, these
permutations yield girths $g=6,8,10,12$, respectively, -- i.e.,
$g=2\ell$. It is clear that the girth of these graphs is upper
bounded by $2\ell$. What is interesting is that 
there exist permutations $\pi,\tau,\tau',\tau''$ that achieve
this upper bound when $\ell \le 6$.
However, when extending this particular construction to
$\ell=7$ layers, there are no permutations $\pi, \tau,\tau',\tau''$
that yield a girth $g=14$ graph. (This was verified by an
exhaustive computer 
search and computing the girths of the resulting graphs using
MAGMA \cite{maxxu}.) The above algorithm to connect the
nodes in layers $L_{\ell-1}$ and $L_{\ell-1}'$ is rather
restrictive, and we need to examine other connection algorithms
that may possibly yield a girth 14 bipartite graph.  However, the smallest known $3$-regular graph with girth
$14$ has 384 vertices \cite{bi98}. For $\ell=7$, the graph of
the Type I-A construction has a total of 380 nodes (i.e., 190
variable nodes and 190 constraint nodes), and there are
permutations $\pi, \tau, \tau', $ and $\tau''$, that only result in a
girth 12 (bipartite) graph.  


{\tiny
\begin{table}
\begin{center}
\begin{tabular}{|c|c|c|c|c|}
\hline
Girth&$g = 6$&$g = 8$&$g = 10$&$g = 12$\\
\hline
$\pi$&(0)(1)&(0)(2)(1,3)&(0)(2)(4)(6)(1,5)(3,7)&(0)(4)(8)(12)(2,6)(10,14)(1,9)(3,15)(5,13)(7,11)\\
$\tau$&$=\pi$&$=\pi$&(0)(2)(4)(6)(1,7)(3,5)&(0)(4)(8)(12)(2,6)(10,14)(1,13)(3,11)(5,9)(7,15)\\
$\tau'$&$=\pi$&$=\pi$&$=\tau$&(0,8)(4,12)(2,14)(6,10)(1,5)(3)(7)(9,13)(11)(15)\\
$\tau''$&$=\pi$&(0,2)(1)(3)&(0,4)(2,6)(1,3)(5,7)&(0,2,4,6)(8,10,12,14)(1,15,5,11)(3,9,7,13)\\
\hline
\end{tabular}
\caption{ Permutations for Type I-A construction.}
\label{table_typeIapermutations}
\end{center}
\end{table}
}


When $\ell =3,5$, the minimum distance of the resulting
code meets the tree bound, and hence, $d_{\min} = w_{\min}$. When
$\ell =4,6$, the minimum distance $d_{\min}$ is strictly larger than the
tree bound; in fact, $d_{\min}$ is more than the tree-bound by
$2$. However,  $w_{\min}=d_{\min}$ for $\ell =4,6$ as well.

\begin{remark}
The Type I-A LDPC codes have  $d_{\min}=w_{\min} =T(d,2\ell)$,  for $\ell=3,5$,
and $d_{\min}=w_{\min}=2+T(d,2\ell)$,  for $\ell=4,6$. 
\label{type1A-dmin-thm}
\end{remark} \vspace{0.1in}


\subsection{Type I-B}
Figure \ref{type1Bd4g6_graph} provides a specific example of a Type I-B 
LDPC constraint graph with $d=4=2^2$.
For $d = p^s$, a prime power, the Type I-B construction yields a $d$-regular 
LDPC constraint graph having $1+d+d(d-1)$ variable and constraint nodes, 
and girth at least $6$. The tree $T$ has 3 layers $L_0,L_1,$ and
$L_2$. The tree is reflected to yield another tree $T'$ and the
variable and constraint nodes in $T'$ are interchanged. Let
$\alpha$ be a primitive element in the field $GF(p^s)$. (Note
that $GF(p^s)$ is the set
$\{0,1,\alpha,\alpha^2,\dots,\alpha^{p^s-2}\}$.) The layer $L_1$ 
(resp., $L_1'$) contains $p^s$ constraint nodes labeled $(0)_c,
(1)_c,(\alpha)_c,\dots,(\alpha^{p^s-2})_c$ (resp., variable nodes
labeled $(0)',(1)',(\alpha)',\dots,(\alpha^{p^s-2})'$). The layer $L_2$ 
(resp., $L_2'$) is composed of $p^s$ sets $\{S_i\}_{i=0,1,\alpha,\alpha^2,\dots,\alpha^{p^s-2}}$ 
of $p^s - 1$ variable (resp., constraint) nodes in each set. Note
that we index the sets by an element of the field $GF(p^s)$. Each 
set $S_i$ corresponds to the children of one of the branches of
the root node. (The `$'$' in the labeling refers to nodes in the
tree $T'$ and the subscript `$c$' refers to constraint nodes.) Let  
$S_i$ (resp., $S_i'$) contain the variable  nodes 
$(i,1),(i,\alpha),\ldots,(i,\alpha^{p^s-2})$ (resp., constraint nodes
$(i,1)'_c,(i,\alpha)'_c,\ldots,(i,\alpha^{p^s-2})'_c$). To use MOLS of
order $p^s$ in the connection algorithm, an imaginary node,
variable node $(i,0)$  (resp., constraint node $(i,0)'_c$) is 
temporarily introduced into each set $S_i$ (resp, $S_i'$). The connection 
algorithm proceeds as follows:

\begin{enumerate}
\item For $i=0,1,\alpha,\dots,\alpha^{p^s-2}$ and
  $j=0,1,\alpha,\dots,\alpha^{p^s-2}$, connect the variable node
  $(i,j)$ in layer $L_2$ to the constraint nodes 
\[{\large (0,j+i\cdot 0)'_c,\ (1,j+i\cdot  1)'_c,\ \ldots,\
  (\alpha^{p^s-2},j+i\cdot \alpha^{p^s-2})'_c}\]
in layer $L'_2$. (Observe that in these connections, every variable node in the
set $S_i$ is mapped to
exactly one constraint node in each set $S'_k$, for
$k=0,1,\alpha,\dots,\alpha^{p^s-2}$, using the array $M_i$ defined in Section 2.B.)

\item Delete all imaginary nodes $\{(i,0),(i,0)'_c\}_{i=0,1,\dots,\alpha^{p^s-2}}$ and 
the edges incident on them.
\item For $i = 1,\ldots,\alpha^{p^s-2},$ delete the edge
  connecting variable node $(0,i)$ to constraint node $(0,i)'_c$.
\end{enumerate}

The resulting $d$-regular constraint graph represents the Type I-B LDPC 
code.\\ \vspace{0.1in}

%

The Type I-B algorithm yields LDPC codes having a wide range of rates and 
blocklengths that are comparable to, but different from, the 
two-dimensional LDPC codes from 
finite Euclidean geometries \cite{ko01,li05}. The Type I-B LDPC
codes are $p^s$-regular with girth at least six, blocklength $N=p^{2s}+1$,
and distance $d_{\min}\ge p^s+1$. For degrees of the form 
$d=2^s$, the resulting binary Type I-B LDPC codes have very good rates, above
0.5, and perform well with iterative decoding. (See Table IV.) \\ \vspace{0.1in}

\begin{thm}
{\em The Type I-B LDPC constraint graphs have a girth of at least six. }
\label{type1B-girth-thm}
\end{thm}

\begin{proof}
We need to show that there are no 4-cycles in the
graph. By construction, it is clear that there are no 4-cycles that involve the
nodes in layers $L_0$, $L_0'$, $L_1$, and $L_1'$. This is because
no two nodes, say, variable nodes $(i,j)$ and $(i,k)$ in a particular class $S_i$ are connected to the
same node $(s,t)'_c$ in some class $S_s'$; otherwise,
 it would mean that $t=j+i\cdot s=k+i\cdot s$. But this is only true for $j=k$.  Therefore,
suppose there is a 4-cycle in the graph, then let us assume that
variable nodes $(i,j)$ and $(s,t)$, for $s\ne i$, are each connected to
constraint nodes $(a,b)'_c$ and $(e,f)'_c$. By construction, this
means that $b=j+i\cdot a=t+s\cdot a$ and $f=j+i\cdot e=t+s\cdot
e$. However then $j-t=(s-i)\cdot a=(s-a)\cdot e$,
thereby implying that $a=e$. When $a=e$, we also have $b=j+i\cdot
a=j+i\cdot e=f$. Thus, $(a,b)'_c=(e,f)'_c$. Therefore, there are no
4-cycles in the Type I-B LDPC graphs. \\
\end{proof}
\vspace{0.1in}

\begin{thm}
{\em The Type I-B LDPC constraint graphs with degree $d=p^s$ and
  girth $g\ge 6$
 have 
\[\begin{array}{cc}
2(p^s-1)\ge d_{\min}\ge w_{\min} \ge T(p^s,6)=1+p^s & \mbox{ for } p>2,\\ 
2(p^s)+1\ge d_{\min}\ge w_{\min} \ge T(p^s,6)=1+p^s & \mbox{ for } p=2.\end{array}
\] 
}
\label{type1B-dmin-thm}
\end{thm}

\begin{proof}
When $p$ is an odd prime, the assertion follows
immediately. Consider the following active variable nodes to be
part of a codeword: variable nodes $(0,1),
(0,\alpha),\dots, (0,\alpha^{p^s-2})$ in $S_0$, and all but the
first variable node in the middle layer $L_1'$ of the reflected
tree $T'$: i.e., variable
nodes $(1)', (\alpha)', (\alpha^2)', \dots, (\alpha^{p^s-2})'$ in
$L_1'$. Clearly all the constraints in $L_2'$ are either connected to none or
exactly two of these active variable nodes. The root node in $T'$
is connected to $p^s-1$ (an even number) active variable nodes
and the first constraint node in $L_1$ of $T$ is also connected to
$p^s-1$ active variable nodes. Hence, these
$2(p^s-1)$ active variable nodes form a codeword. This fact along with
Theorems~\ref{thm1} and \ref{type1B-girth-thm} prove that $2(p^s-1)\ge d_{\min}\ge
w_{\min}\ge T(p^s,6)=1+p^s$.\\

When $p=2$,  consider the following active variable nodes to be part of a codeword: 
the root node, variable nodes $(0,1),(0,\alpha),\dots,(0,\alpha^{p^s-2})$ in $S_0$, 
variable node $(\alpha^i,\alpha^i)$ from $S_{\alpha^i}$, for $i=0,1,2,\dots,p^s-2$, 
and the first two variable nodes in the middle layer of $T'$ (i.e., variable nodes $(0)'$,$(1)'$).  
Since $p=2$, $p^s-1$ is odd. We need to 
show that all the constraints are satisfied for this choice of active variable nodes.
Each constraint node in the layer $L_1$ of $T$ has 
an even number of active variable node neighbors: $(0)_c$ has $p^s$ active neighbors, and
$(i)_c$, for $i=1,\alpha,\dots,\alpha^{p^s-2}$, has two, the root node and 
variable node $(i,i)$. It remains to check the constraint nodes in $T'$.

In order to examine the constraints in layer $L_2'$ of $T'$, observe that
the variable node $(0,\alpha^j)$, for $j=1,2,\dots,p^s-2$, is connected to constraint nodes 
\[ (1,\alpha^j)'_c, (\alpha,\alpha^j)'_c,\dots,(\alpha^{p^s-2},\alpha^j)'_c,\] 
and the variable node $(\alpha^i,\alpha^i)$, for
$i=0,1,2,\dots,p^s-2$, is connected to constraint nodes
\[(0,\alpha^i)'_c, (1,\alpha^i+\alpha^i)'_c, (\alpha,\alpha^i+\alpha^{i+1})'_c,\dots,(\alpha^k,\alpha^{i+k})'_c,\dots,
(\alpha^{p^s-2},\alpha^{i+p^s-2})'_c\] 

Therefore, the constraint nodes $(0,\alpha^j)'_c$, for $j=1,2,\dots,p^s-2$, in $S_0'$ of $L_2'$ 
are connected to exactly one active variable node from layer $L_2$, i.e., variable node $(\alpha^j,\alpha^j)$; the other active variable node neighbor is variable node $(0)'$
in the middle layer of $T'$. Thus, all constraints in $S_0'$ are satisfied.

The constraint nodes $(1,\alpha^j)'_c$, for $j=1,2,\dots,p^s-2$, in $S_1'$ are each connected
to exactly one active variable node from $L_2$, i.e., variable node $(0,\alpha^j)$ from $S_0$.
This is because, all the remaining active variable nodes in $L_2$, $(\alpha^i,\alpha^i)$ connect to
the imaginary node $(1,0)'_c$ in $S_1'$ (since $(1,\alpha^i+\alpha^i)'_c=(1,0)'_c$ when the characteristic of
the field $GF(p^s)$ is $p=2$). Thus, all constraint nodes in $S_1'$ have two active variable node neighbors, the other active neighbor being the variable node $(1)'$ in the  middle layer of $T'$. 

Now, let us consider the constraint nodes in $S_{\alpha^k}'$, for $k=1,2,\dots,p^s-2$. The active variable nodes 
$(\alpha^i,\alpha^i)$, for $i=0,1,2,\dots,p^s-2$, are connected
to the following constraint nodes
\[(\alpha^k,1+\alpha^k)'_c,
(\alpha^k,\alpha+\alpha^{k+1})'_c,\dots,(\alpha^k,\alpha^{p^s-2}+\alpha^{p^s-2+k})'_c,\]
respectively, in class $S_{\alpha^k}'$.
Since $\alpha^r+\alpha^{k+r} \ne \alpha^t+\alpha^{k+t}$ for $r\ne t$, the variable nodes 
$(\alpha^i,\alpha^i)$, for $i=0,1,2,\dots,p^s-2$, connect to distinct nodes in $S_{\alpha^k}'$.
Hence, each constraint node in $S_{\alpha^k}'$ has exactly two
active variable node neighbors -- one from $S_0$ and the other from the set 
$\{(\alpha^i,\alpha^i)|\ i=0,1,2,\dots,p^s-2\}$. 

Last, we note that the root (constraint) node in $T'$ is connected to two active variable nodes,
$(0)'$ and $(1)'$. The total number of active variable nodes is $1+(p^s-1)+(p^s-1)+2=2p^s+1$. 
This proves that the set of $2p^s+1$ active variable nodes forms a codeword, thereby proving the desired bound. \\
\end{proof}
\vspace{0.1in}

When $p>2$, the upper bound $2(p^s-1)$ on minimum distance
$d_{\min}$ (and possibly also $w_{\min}$) was met among all the
cases of the Type I-B construction  we examined. We conjecture that
in fact $d_{\min}=2(p^s-1)$ for the Type I-B LDPC codes of degree
$d=p^s$ when $p>2$. Since $w_{\min}$ is lower bounded by
$1+p^s$, we have that $w_{\min}$ is close, if not equal, to $d_{\min}$.\\
\vspace{0.1in}

\subsection{$p$-ary LDPC codes}
Let $H$ be a parity check matrix representing a $p$-ary
LDPC code $\mathcal{C}$. The corresponding LDPC constraint graph $G$
that represents $H$ is an incidence graph of the parity check matrix as in
the binary case. However, each edge of $G$ is now assigned a
weight which is the value of the corresponding non-zero entry in
$H$. (In \cite{da98,da99t}, LDPC codes over $GF(q)$ are
considered for transmission over binary modulated
channels, whereas in \cite{sr05}, LDPC codes over integer rings
are considered for higher-order modulation signal sets.)

For convenience, we consider the special case wherein each of
these edge weights are equal to one. This is the case when the
parity check matrix has only zeros and ones. Furthermore,
whenever the LDPC graphs have edge weights of unity for all the
edges, we refer to such a graph as a binary LDPC constraint graph
representing a $p$-ary LDPC code $\mathcal{C}$.

We first show that if the LDPC graph
corresponding to $H$ is $d$-left (variable-node) regular, then the same tree
bound of Theorem~\ref{thm1}  holds. That is,\\

\begin{lem}
{\em If $G$ is a $d$-left regular bipartite  LDPC constraint graph with girth $g$
and represents a $p$-ary LDPC code $\mathcal{C}$. Then, the minimum distance of
the $p$-ary LDPC code $\mathcal{C}$ is lower bounded as 
 \[d_{\min} \ge T(d,g). \]
}
\label{q_ary_dmin}
\end{lem}

\begin{proof}
The proof is essentially the same as in the binary
case. Enumerate the graph as a tree starting at an arbitrary
variable node. Furthermore, assume that a codeword in
$\mathcal{C}$ contains the root node in its support. The root
variable node (at layer $L_0$ of the tree) connects to $d$
constraint nodes in the next layer (layer $L_1$)
of the tree. These constraint nodes are each connected
to some sets of variable nodes in layer $L_2$, and so on. Since
the graph has girth $g$, the nodes enumerated up to layer
$L_{\frac{g-6}{2}}$ when $\frac{g}{2}$ is odd (respectively,
$L_{\frac{g-4}{2}}$ when $\frac{g}{2}$ is even) are all
distinct. Since the root node belongs to a codeword, say ${\bf c}$, it assumes a
non-zero value in ${\bf c}$. Since the constraints must be
satisfied at the nodes in layer $L_1$, at least one node in Layer
$L_2$ for each constraint node in $L_1$ must assume a
non-zero value in ${\bf c}$. (This is under the assumption that
an edge weight 
times a (non-zero) value, assigned to the corresponding variable node, is not
zero in the code alphabet. Since we have chosen the edge weights to be unity, such a
case will not arise here. But also more generally, such cases will not arise when the
alphabet and the arithmetic operations are that of a finite
field. However, when working over other structures, such as
finite integer rings and more general groups, such cases could arise.)

Under the above assumption, that there are at least $d$ variable
nodes (i.e., at least one for each node in layer $L_1$) in layer
$L_2$ that are non-zero in ${\bf c}$. Continuing this argument,
it is easy to see that the number of non-zero components in ${\bf c}$ is at least
$1+d+d(d-1)+\dots+d(d-1)^{\frac{g-6}{4}}$ when $\frac{g}{2}$ is
odd, and $1+d+d(d-1)+\dots+d(d-1)^{\frac{g-8}{4}}
+(d-1)^{\frac{g-4}{4}}$ when $\frac{g}{2}$ is even. Thus, the
desired lower bound holds.\\
\end{proof}
\vspace{0.1in}

We note here that in general this lower bound is not met and
typically $p$-ary LDPC codes that have the above graph
representation have minimum distances larger than the above lower
bound.

Recall from \cite{ko03p,ke05u} that a pseudocodeword of an LDPC constraint graph $G$ is
a valid codeword in some finite cover of $G$. To define a
pseudocodeword for a $p$-ary LDPC code, we will restrict the
discussion to LDPC constraint graphs that have edge weights of
unity among all their edges -- in other words, binary LDPC constraint graphs
that represent $p$-ary LDPC codes.  A finite cover of a graph
is defined in a natural way as in \cite{ko03p} wherein all edges in the
finite cover also have an edge weight of unity. For the rest of
this section, let $G$ be a LDPC constraint graph of a $p$-ary LDPC
code $\mathcal{C}$ of block length $n$, and let the weights on every edge of $G$ be
unity. We define a pseudocodeword $F$ of $G$ as a $n\times p$ matrix of the form
\[ F= \left[\begin{array}{ccccc} 
f_{0,0}&f_{0,1}& f_{0,2}&\dots&f_{0,p-1}\\
f_{1,0}&f_{1,1}& f_{1,2}&\dots&f_{1,p-1}\\
\vdots&\vdots&\vdots&\vdots&\vdots\\
f_{n-1,0}&f_{n-1,1}& f_{n-1,2}&\dots&f_{n-1,p-1}
\end{array}\right] , \]
where the pseudocodeword $F$ forms a valid codeword
$\hat{\bf c}$ in a finite cover $\hat{G}$ of $G$ and $f_{i,j}$ is
the fraction of variable nodes in the $i^{th}$ variable cloud, for
$0\le i\le n-1$, of
$\hat{G}$ that have the assignment (or, value) equal to $j$, for
$0\le j\le p-1$, in $\hat{\bf c}$. 

A $p$-ary symmetric channel is shown in
Figure~\ref{q_ary_sym}. The input and the 
output of the channel are random variables belonging to a $p$-ary
alphabet that can be denoted as $\{0,1,2,\dots,p-1\}$. An error
occurs with probability $\epsilon$, which is parameterized by the
channel, and in the case of an error,
it is equally probable for an input symbol to be altered to any one of the remaining
symbols.\\

Following the definition of pseudoweight for the binary symmetric
channel \cite{fo01in}, we provide the following definition for the
weight of a pseudocodeword on the $p$-ary symmetric channel.
For a pseudocodeword $ F$, let $ F'$ be the sub-matrix
obtained by removing the first column in $F$. (Note that the
first column in $F$ contains the entries $f_{0,0}, f_{1,0},
f_{2,0}, \dots,f_{n-1,0}$.) Then the weight of a pseudocodeword
$F$ on the $p$-ary symmetric channel is defined as follows.\\

\begin{definition} Let $e$ be a number such that the sum of the $e$ largest components in
the matrix $F'$, say,
$f_{i_1,j_1},f_{i_2,j_2},\dots,f_{i_e,j_e}$, exceeds $\sum_{i\ne i_1,i_2,\dots,i_e}(1-f_{i,0})$.
Then the weight of $F$ on the $p$-ary symmetric channel is
defined as 
\[ w_{PSC}(F) =\left\{ \begin{array}{cc}
           2e, & \mbox{if } f_{i_1,j_1}+\dots +f_{i_e,j_e} =\sum_{i\ne i_1,i_2,\dots,i_e}(1-f_{i,0}),\\
2e-1, & \mbox{if } f_{i_1,j_1}+\dots +f_{i_e,j_e} > \sum_{i\ne i_1,i_2,\dots,i_e}(1-f_{i,0}).
\end{array} \right  . \vspace{0.1in}
\] 
\end{definition}
\vspace{0.1in}

Note that in the above definition, none of the $j_k$'s, for
$k=1,2,\dots,e$, are equal to zero, and all the $i_k$'s, for
$k=1,2,\dots,e$, are distinct. That is, we choose at most one
component in every row of $F'$ when picking the $e$
largest components. (See the appendix for an explanation on the
above definition of ``weight''.)

Observe that for a codeword, the above weight definition reduces to the
Hamming weight. If $F$ represents a codeword ${\bf c}$, then exactly
$w=wt_H({\bf c})$, the Hamming weight of ${\bf c}$, rows in
$F'$ contain the entry $1$ in some column, and the remaining
entries in $F'$ are zero. Furthermore, the matrix $F$
has the entry $0$ in the first column of these $w$ rows and has the
entry $1$ in the first column of the remaining rows. Therefore,
from the weight definition of $F$, 
$e=\frac{w}{2}$ and the weight of $F$ is $2e=w$.

We define the $p$-ary minimum pseudocodeword weight of $G$ (or, minimum
pseudoweight) as in the binary case, i.e., as the minimum weight
of a pseudocodeword among all finite covers of $G$, and denote
this as $w_{\min}(G)$ or $w_{\min}$ when it is clear that we are
referring to the graph $G$. \\

\begin{lem}
{\em Let $G$ be a $d$-left regular bipartite graph with girth $g$ that
represents a $p$-ary LDPC code $\mathcal{C}$. Then the minimum
pseudocodeword weight 
$w_{\min}$ on the $p$-ary symmetric channel is lower bounded as
\[w_{\min} \ge T(d,g)=\left\{\begin{array}{cc} 
(1+d + d(d-1) + d(d-1)^2+\ldots+ d(d-1)^{\frac{g-6}{4}}), & \frac{g}{2}\mbox{ odd },\\
(1+ d+d(d-1)+\ldots+d(d-1)^{\frac{g-8}{4}} + (d-1)^{\frac{g-4}{4}}),& \frac{g}{2}\mbox{ even }.
\end{array}\right . \]
}
\label{q_ary_wmin}
\end{lem}
\vspace{0.1in}

The proof of this result is moved to the appendix. We note that,
in general, this bound is rather loose. (The inequality in
(\ref{eqn_spc}), in the proof of Lemma~\ref{q_ary_wmin}, is
typically not tight.)  Moreover, we expect that $p$-ary LDPC
codes to have larger minimum pseudocodeword weights than
{\em corresponding binary LDPC codes}. By {\em corresponding binary LDPC
codes}, we mean the codes obtained by interpreting the given LDPC
constraint graph as one representing a binary LDPC code.\\

\subsection{$p$-ary Type I-B LDPC codes}

\begin{thm}
{\em For degree $d=p^s$, the resulting Type I-B LDPC constraint graphs
of girth $g$ that represent $p$-ary LDPC codes have minimum distance and
minimum pseudocodeword weight \[2p^s+1\ge d_{\min}\ge w_{\min} \ge T(d,g).\] 
}
\label{type1B-Pary-dmin-thm}
\end{thm}

\begin{proof}
Consider as active variable nodes the root node, all the
variable nodes in $S_0$, the variable nodes
$(\alpha^i,\alpha^i)$, for $i=0,1,2,\dots,p^s-2$, the first
variable node $(0)'$ in the middle layer of $T'$ and one other
variable node $(y)'$, that we will ascertain later, in the middle
layer of $T'$. 

Since the code is $p$-ary (and $p>2$), assign the value 1 to the
root variable node and to all the active variable nodes in $S_0$. Assign
the value $p-1$ to
the remaining active variable nodes in $L_2$, (i.e., nodes
$(\alpha^i,\alpha^i)$, $i=0,1,2,\dots,p^s-2$). Assign the value
$1$ for the variable node $(0)'$ in the middle layer of $T'$ and
assign the value $p-1$ for the variable node $(y)'$ in the middle
layer of $T'$. We choose $y$ in the following manner:

The variable nodes $(\alpha^i,\alpha^i)$, for
$i=0,1,2,\dots,p^s-2$, are connected to the following constraint nodes 
\[(\alpha^k, 1+\alpha^k)'_c,
(\alpha^k,\alpha+\alpha^{k+1})'_c,\dots,(\alpha^k,\alpha^j+\alpha^{k+j})'_c,\dots,(\alpha^k,\alpha^{p^s-2}+\alpha^{k+p^s-2})'_c,\]
respectively, in class $S_{\alpha^k}$. 
Either, the above set of constraint nodes are all distinct, or
they are all equal to $(\alpha^k,0)'_c$. This is because,
$\alpha^r+\alpha^{r+k}=\alpha^t+\alpha^{t+k}$ if and only if
either, $r=t$ or $1+\alpha^k=0$. So there is only one $k \in
\{0,1,2,\dots,p^s-2\}$, for which $1+\alpha^k=0$, and for that
value of $k$, we set $y=\alpha^k$. 

From the proof of Theorem~\ref{type1B-dmin-thm} and the above
assignment, it is easily verified  that each constraint node has value zero when the
sum of the incoming active nodes is take modulo $p$.  Thus, the set of $2p^s+1$ 
active variable nodes forms a codeword, and therefore $d_{\min}\le 2p^s+1$. Hence, from
Lemmas~\ref{q_ary_dmin} and \ref{q_ary_wmin}, we have $T(d,6)\le w_{\min}\le d_{\min} 
\le 2p^s+1$.\\
\end{proof}
\vspace{0.1in}

It is also observed that if the codes resulting from the Type
I-B construction are treated as $p$-ary codes rather than binary
codes when the corresponding degree in the LDPC graph is $d=p^s$,
then the rates obtained are also $> 0.5$. (See Table IV). We also believe
that the minimum pseudocodeword weights (on the $p$-ary symmetric
channel) are much closer to the minimum distances for these $p$-ary LDPC codes.\\
\vspace{0.1in}

\section{Tree-based Construction: Type II}
In the Type II construction, first a $d$-regular tree $T$ of
alternating variable and constraint node layers is enumerated
from a root variable node (layer $ L_0$) for $\ell$ layers $L_0,L_1,\dots,L_{\ell-1}$, as in
Type I. The tree $T$ is not reflected; rather, a single layer of
$(d-1)^{\ell-1}$ nodes is added to form layer $L_{\ell}$. If
$\ell$ is odd (resp., even), this layer is composed of constraint
(resp., variable) nodes. The union of $T$ and $L_{\ell}$,  along
with edges connecting the nodes in layers $L_{\ell-1}$ and
$L_{\ell}$ according to a connection algorithm that is described
next, comprise the graph representing a Type II LDPC code. We 
present the  connection scheme that is used for this Type II
model, and discuss the resulting codes. First, we state this
rather simple observation without proof: 

{\em The girth $g$ of a Type II LDPC graph for $\ell$ layers is at
most $2\ell$.}\\

\noindent  The connection algorithm for $\ell = 3$ and $\ell = 4$,
wherein this upper bound on girth is in fact achieved, is as follows.\\

\subsection{ $\ell = 3$}
Figure \ref{type2g6d4_graph} provides an example of a Type II
LDPC constraint graph for $\ell=3$ layers, with degree $d=4=3+1$ and girth $g=6$.
For $d=p^s+1$, where $p$ is prime and $s$ a positive
integer, a $d$-regular tree is enumerated from a root (variable) node for
$\ell=3$ layers $L_0,L_1, L_2$. Let $\alpha$ be a primitive
element in the field $GF(p^s)$. 
The $d$ constraint nodes in $L_1$ are labeled
$(x)_c,(0)_c,(1)_c,(\alpha)_c,\dots,(\alpha^{p^s-2})_c$ to represent
the $d$ branches stemming from the root node. Note that the first
constraint node is denoted as $(x)_c$ and the remaining
constraint nodes are indexed
by the elements of the field $GF(p^s)$.  The $d(d-1)$ variable
nodes in the third layer $L_2$ are labeled as follows:
the variable nodes descending from constraint node $(x)_c$
form the class $S_x$ and are labeled $(x,0), (x,1), \dots,(x,\alpha^{p^s-2})$, and the
variable nodes descending from constraint node $(i)_c$, for
$i=0,1,\alpha,\dots,\alpha^{p^s-2}$, form the class $S_i$ and are labeled $(i,0), (i,1), \dots,(i,\alpha^{p^s-2})$.

A final layer $L_{\ell}=L_3$ of $(d-1)^{\ell-1}=p^{2s}$ constraint nodes
is added. The $p^{2s}$ constraint nodes in $L_{3}$
are labeled $(0,0)'_c, (0,1)'_c,\dots, (0,\alpha^{p^s-2})'_c$, 
$(1,0)'_c$, $(1,1)'_c$, $\dots$, $(1,\alpha^{p^s-2})'_c$, $\dots$,
$(\alpha^{p^s-2},0)'_c$, $(\alpha^{p^s-2},1)'_c$, $\dots$,
$(\alpha^{p^s-2},\alpha^{p^s-2})'_c$. (Note that the `$'$' in the
labeling refers to nodes in that are not in the tree $T$ and the subscript `$c$' refers to constraint nodes.)

\begin{enumerate}
\item By this labeling, the constraint nodes in $L_3$ are grouped into $d-1=p^s$  classes of $d-1=p^s$ nodes in each class. Similarly, the
  variable nodes in $L_2$ are grouped into $d=p^s+1$ classes of
  $d-1=p^s$ nodes in each class. (That is, the  $i^{th}$ class of
  constraint nodes is $S_i'=\{(i,0)'_c,(i,1)'_c,\dots,(i,\alpha^{p^s-2})'_c\}$.)
\item The variable nodes descending from constraint node $(x)_c$ are
  connected to the constraint nodes in $L_3$ as follows. Connect
  the variable node $(x,i)$, for $i=0,1,\dots,\alpha^{p^s-2}$, to
  the constraint nodes \[ (i,0)'_c,\ (i,1)'_c,\ \dots,\ (i,\alpha^{p^s-2})'_c.\]

\item The remaining variable nodes in layer
  $L_2$ are connected to the nodes in $L_3$ as  follows: Connect the variable node $(i,j)$, for
  $i=0,1,\alpha,\dots, \alpha^{p^s-2}$,
  $j=0,1,\dots,\alpha^{p^s-2}$, to the constraint nodes
\[(0,j+i\cdot 0)'_c,\ (1,j+i\cdot 1)'_c,\ (\alpha,j+i\cdot \alpha)'_c,\
\dots,\ (\alpha^{p^s-2},j+i\cdot \alpha^{p^s-2})'_c .\]
Observe that in these connections, each variable node $(i,j)$ is
connected to exactly one constraint node within each class, using the
array $M_i$ defined in Section 2.B.\\
\end{enumerate}\vspace{-0in}

In the example illustrated in Figure \ref{type2g6d4_graph}, the
arrays used for constructing the Type II LDPC constraint graph are\footnote{Note that
  in this example,  $GF(3)=\{0,1,2\}$, `$2$' being the primitive
  element of the field.} 
\[\tiny 
\hspace{-0in} M_{0}=\left[\begin{array}{ccc}
0&0&0\\
1&1&1\\
2&2&2
\end{array}\right], \ M_{1}=  \left[\begin{array}{ccc}
0&1&2\\
1&2&0\\
2&0&1
\end{array}\right], \ M_{2}=\left[\begin{array}{ccc}
0&2&1\\
1&0&2\\
2&1&0
\end{array}\right].\]

The ratio of minimum distance to blocklength of the resulting
codes is at least $\frac{2+p^s}{1+p^s+p^{2s}}$, and the girth is
six. For degrees $d$ of the form $d=2^s+1$, the tree bound of
Theorem~\ref{thm1} on minimum distance and minimum pseudocodeword
weight \cite{ta81a,ke04p} is met, i.e.,
$d_{\min}=w_{\min}=2+2^s$, for the Type II, $\ell=3$, LDPC
codes. For $p>2$, the resulting binary LDPC codes are repetition codes of the
form $[n,1,n]$, i.e., $d_{\min}=n=1+p^s+p^{2s}$ and the rate is $\frac{1}{n}$. However, if we
interpret the Type II $\ell=3$ graphs, that have degree $d=p^s+1$, as
the LDPC constraint graph of a $p$-ary LDPC code, then the rates
of the resulting codes are very good and the minimum distances
come close to (but are not equal to) the tree bound in
Lemma~\ref{q_ary_dmin}. (See also \cite{ru67}.) In summary, we state the following results:\\

\begin{itemize}
\item The rate of a p-ary Type II, $\ell=3$ LDPC code is
$\frac{p^{2s}+p^s-\frac{p^s(p+1)^s}{2^s}}{p^{2s}+p^s+1}$ \cite{mac}.
\item The rate of a binary Type II, $\ell = 3$ LDPC code is $\frac{1}{n}$ for $p > 2$.
\end{itemize}

Note that binary codes with $p = 2$ are a special case of $p$-ary LDPC codes.  Moreover, the rate
expression for $p$-ary LDPC codes is meaningful for a wide variety of $p$'s and $s$'s. The rate
expression for binary codes with $p>2$ can be seen by observing that any $t$ rows of the
corresponding parity-check matrix $H$ is linearly independent if $t < n$. Since the parity-check
matrix is equivalent to one obtainable from cyclic difference sets, this can be proven by showing
that for any
$t < n$, there exists a set of
$t$ consecutive positions in the first row of
$H$ that has an odd number of ones.\\

\subsection{Relation to finite geometry codes}
The codes that result from this $\ell = 3$ construction correspond to the two-dimensional projective-geometry-based LDPC (PG LDPC) codes of \cite{li05}. We state the equivalence of the tree construction and the finite projective geometry based LDPC codes in the following.

\begin{thm}
{\em The LDPC constraint graph obtained from the Type II $\ell=3$ tree
construction for degree $d=p^s+1$ is equivalent to the incidence
graph of the finite projective plane over the field $GF(p^s)$.
}
\end{thm} 
\vspace{0.1in}

It has been proved by Bose \cite{bo38} that a finite projective
plane (in other words, a two dimensional finite projective
geometry) of order $m$ exists if and only if a complete family of
orthogonal $m\times m$ Latin squares exists. The proof of this
result, as presented in \cite{ro84b}, gives a constructive
algorithm to design a finite projective plane of order $m$ from a
complete family of $m\times m$ mutually orthogonal Latin squares
(MOLS). It is well known that a complete family of mutually
orthogonal Latin squares exists when $m=p^s$, a power of a prime,
and we have described one such family in Section 2. Hence, the
constructive algorithm in \cite{ro84b} generates the incidence
graph of the projective plane $PG(2,p^s)$  from the set of
$p^s-1$ MOLS of order $p^s$.  The only remaining step is to verify that the incidence matrix of points over lines of this projective plane is the same as the parity check matrix of variable nodes over constraint nodes of the tree-based LDPC constraint graph of the tree construction. This step is easy to verify as the constructive algorithm in \cite{ro84b} is analogous to the tree construction presented in this paper.
\\

The Type II $\ell=3$ graphs therefore correspond to the two-dimensional projective-geometry-based LDPC codes of
\cite{ko01}. With a little modification of the Type II construction, we can also obtain the two-dimensional Euclidean-geometry-based LDPC codes of \cite{ko01,li05}. Since a two-dimensional Euclidean geometry may be obtained by deleting certain points and line(s) of a two-dimensional projective geometry, the  graph of a two-dimensional EG-LDPC code \cite{li05} may be obtained by performing the following operations on the Type II, $\ell = 3$, graph:
\begin{enumerate}
\item In the tree $T$, the root node along with its neighbors, i.e., the constraint nodes in layer $L_1$, are deleted.
\item Consequently, the edges from the constraint nodes $(x)_c,(0)_c,(1)_c,\ldots,(\alpha^{p^s-2})_c$ to layer $L_2$ are also deleted.

\item At this stage, the remaining variable nodes have degree $p^s$, and 
the remaining constraint nodes have degree $p^s+1$. Now, a constraint node 
from layer $L_3$ is chosen, say, constraint node $(0,0)'_c$. This node and 
its neighboring variable nodes and the edges incident on them are deleted. 
Doing so removes exactly one variable node from each class of $L_2$, and 
the degrees of the remaining constraint nodes in $L_3$ are lessened by 
one. Thus, the resulting graph is now $p^s$-regular with a girth of six, 
has $p^{2s}-1$ constraint and variable nodes, and corresponds to the 
two-dimensional Euclidean-geometry-based LDPC code $EG(2,p^s)$ of 
\cite{li05}.\\

\end{enumerate}

\begin{thm}
{\em The Type II $\ell=3$ LDPC constraint graphs have girth $g=6$ and
  diameter $\delta=3$. 
}
\label{type2ell3-girth-thm}
\end{thm}

\begin{proof}
We need to show is that there are no 4-cycles in the
graph. As in the proof of Theorem~\ref{type1B-girth-thm}, by construction, there are no 4-cycles that involve the
nodes in layers $L_0$ and $L_1$. This is
because, first, no two variable nodes in the first class $S_x=
\{(x,0),(x,1),\dots,(x,\alpha^{p^s-2})\}$ are connected to the same
constraint node. Next, if two variable nodes, say, $(i,j)$ and
$(i,k)$ in the $i^{th}$ class $S_i$, for some $i\ne x$, are
connected to a  constraint node $(s,t)'_c$, then it would mean that
$t=j+i\cdot s=k+i\cdot s$. But this is only true for
$j=k$. Hence, there is no 4-cycle of the form $(i)_c\rightarrow (i,j)\rightarrow
(s,t)'_c \rightarrow (i,k)\rightarrow (i)$. Therefore,
suppose there is a 4-cycle in the graph, then let us consider two
cases as follows. Case 1) Assume that
variable nodes $(i,j)$ and $(s,t)$, for $i\ne s$ and $i\ne x\ne s$, are each connected to
constraint nodes $(a,b)'_c$ and $(e,f)'_c$. By construction, this
means that $b=j+i\cdot a=t+s\cdot a$ and
$f=j+i\cdot e=t+s\cdot e$. This implies that $j-t=(s-i)\cdot
a=(s-i)\cdot e$, thereby implying that $a=e$. Consequently, we
also have $b=j+i\cdot a= j+i\cdot e=f$. Thus, $(a,b)'_c=(e,f)'_c$.
Case 2) Assume that two variable nodes, one in $S_x$, say,
$(x,j)$, and the other in $S_i$, (for $i\ne x$), say, $(i,k)$, are
connected to constraint nodes $(a,b)'_c$ and $(e,f)'_c$. Then this
would mean that $a=e=j$. But since $(i,k)$ connects to exactly
one constraint node whose first index is $j$, this case
is not possible. Thus, there are no 4-cycles in the Type II-$\ell=3$ LDPC
graphs. 

To show that the girth is exactly six, we see that the following
nodes form  a six-cycle in the graph: the root-node, the first two constraint
nodes $(x)_c$ and $(1)_c$ in layer $L_1$, variable nodes $(x,0)$ and
$(0,0)$ in layer $L_2$, and the constraint node $(0,0)'_c$ in layer $L_3$.

To prove the diameter, we first observe that the root node is at
distance of at most
three from any other node. Similarly, it is also clear that the
nodes in layer $L_1$ are at a distance of at most three from any
other node. Therefore, it is only necessary to show that any two
nodes in layer $L_2$ are at most distance two apart and
similarly show that any two nodes in $L_3$ are at most distance two
apart.   Consider two nodes $(i,j)$ and $(s,t)$ in $L_2$. If
$s=i$, then clearly, there is a path of length two via the parent
node $(i)_c$. If $s\ne i$ and $s\ne x\ne i$, then by the property of a complete
family of orthogonal Latin squares there is a node $(a,b)'_c$ in
$L_3$ such that $b=j+i\cdot a=t+s\cdot a$. This implies
that $(i,j)$ and $(s,t)$ are connected by a distance two path
via $(a,b)'_c$. We can similarly show that if $s\ne i$ and
$i=x$, then the node $(j,t+s\cdot j)'_c$ in $L_3$ connects to both
$(x,j)$ and $(s,t)$. A similar argument shows that any two
nodes in $L_3$ are distance two apart. This completes the proof. \\
\end{proof}
\vspace{0.1in}

\begin{thm}
{\em For degrees $d=2^s+1$, the resulting Type II $\ell=3$ LDPC
constraint graphs have \[d_{\min}=w_{\min}=T(d,6)=2+2^s.\]
For degrees $d=p^s+1$, $p>2$, when the resulting Type II $\ell=3$
LDPC constraint graphs represent $p$-ary linear codes, the
corresponding minimum distance and minimum pseudocodeword weight
satisfy \[T(p^s+1,6)\le w_{\min}\le d_{\min}\le 2p^s.\] 
}
\label{type2ell3-dmin-thm}
\end{thm}

\begin{proof}
Let us first consider the case $p=2$. We will show that the
following set of {\em active} variable nodes in the Type II $\ell=3$ LDPC
constraint graph form a minimum-weight codeword:
the root (variable) node, variable nodes $(x,0)$, $(0,0)$,
$(1,\alpha^{p^s-2})$, $(\alpha,\alpha^{p^s-3})$,
$(\alpha^2,\alpha^{p^s-4})$, $\dots$,
$(\alpha^i,\alpha^{p^s-2-i})$, $\dots$,  $(\alpha^{p^s-2},1)$ in
layer $L_2$.
 
It is clear from this choice that there is exactly one active variable node from each
class in layer $L_2$. Therefore, all the constraint nodes at
layer $L_1$ are satisfied. The constraint nodes in the first
class $S_0'$ of $L_3$ are $(0,0)'_c, (0,1)'_c,\dots,
(0,\alpha^{p^s-2})'_c$. The constraint node $(0,0)'_c$ is connected to $(x,0)$
and $(0,0)$, and the constraint node $(0,\alpha^i)'_c$, for $i=0,1,2,\dots,p^s-2$, is connected to variable nodes
$(x,0)$ and $(\alpha^i,\alpha^{p^s-2-i})$. Thus, all constraint
nodes in $S_0'$ are satisfied. Let us consider the constraint
nodes in class $S_{\alpha^i}'$, for
$i\in\{0,1,2,\dots,p^s-2\}$. The variable node
  $(0,0)$ connects to the constraint node $(\alpha^i,0)'_c$ in $S_{\alpha^i}'$. The variable
  node $(1,\alpha^{p^s-2})$ connects to the constraint node
  $(\alpha^i,\alpha^{p^s-2}+\alpha^i)'_c$ in $S_{\alpha^i}'$,  and in
  general, for $j=0,1,2,\dots,p^s-2$, the variable
  node $(\alpha^j, \alpha^{p^s-2-j})$  connects to the constraint
  node  $(\alpha^i,\alpha^{p^s-2-j}+\alpha^{i+j})'_c$ in $S_{\alpha^i}'$. 
So enumerating all the constraint nodes in $S_{\alpha^i}'$, with
multiplicities, that are connected to an active variable node in $L_2$,
we obtain\\
$(\alpha^i,0)'_c$, $(\alpha^i,\alpha^{p^s-2}+\alpha^i)'_c$,
$(\alpha^i,\alpha^{p^s-3}+\alpha^{i+1})'_c$, $\dots$,
$(\alpha^i,\alpha^{p^s-(p^s-i-1)}+\alpha^{p^s-3})'_c$,
$(\alpha^i,\alpha^{p^s-(p^s-i)}+\alpha^{p^s-2})'_c$, $\dots$.

Simplifying the exponents and rewriting this list, we see that,
when $i$ is odd, the constraint nodes are 

$(\alpha^i,0)'_c$, $(\alpha^i,\alpha^{p^s-2}+\alpha^i)'_c$,
$(\alpha^i,\alpha^{p^s-3}+\alpha^{i+1})'_c$,
$\dots$,
$(\alpha^i,\alpha^{i+1}+\alpha^{p^s-3})'_c$,
$(\alpha^i,\alpha^{i}+\alpha^{p^s-2})'_c$, $(\alpha^i,\alpha^{i-1}+1)'_c$,
$(\alpha^i,\alpha^{i-2}+\alpha)'_c$, $\dots$,
$(\alpha^i,\alpha^{(i-1)/2}+\alpha^{(i-1)/2})'_c$,$\dots$,
$(\alpha^i,\alpha+\alpha^{i-2})'_c$, $(\alpha^i,1+\alpha^{i-1})'_c$.

(When $i$ is even, the constraint nodes are
$(\alpha^i,0)'_c$, $(\alpha^i,\alpha^{p^s-2}+\alpha^i)'_c$,
$(\alpha^i,\alpha^{p^s-3}+\alpha^{i+1})'_c$,
$\dots$,
$(\alpha^i,\alpha^{(p^s-2+i)/2}+\alpha^{(p^s-2+i)/2})'_c$, $\dots$,
$(\alpha^i,\alpha^{i+1}+\alpha^{p^s-3})'_c$,
$(\alpha^i,\alpha^{i}+\alpha^{p^s-2})'_c$, $(\alpha^i,\alpha^{i-1}+1)'_c$,
$(\alpha^i,\alpha^{i-2}+\alpha)'_c$, $\dots$,
$(\alpha^i,\alpha^{i}+\alpha^{i-1})'_c$,
$(\alpha^i,\alpha^{i-1}+\alpha^i)'_c$,$\dots$, $(\alpha^i,1+\alpha^{i-1})'_c$.)
(Note that for $\beta\in GF(p^s)$, $\beta+\beta=0$ when the characteristic
of the field $GF(p^s)$ is two (i.e., $p=2$).)

Observe that each of the constraint nodes in the above list appears
exactly twice. Therefore, each constraint node in the list is
connected to two active variable nodes in $L_2$, and hence, all the
constraints in $S_{\alpha^i}'$ are satisfied. So we have that  the set of
$1+2^s+1$ active variable nodes forms a codeword. Furthermore,
they must form a minimum-weight codeword since $d_{\min}\ge 2+2^s=T(2^s+1,6)$ by
the tree bound of Theorem~\ref{thm1}. This also proves that
$d_{\min}=w_{\min}=T(d,6)$ for $d=2^s+1$.\\

Now let us consider the case $p>2$. The resulting codes are
treated as $p$-ary codes. Consider the following set of active
variable nodes: the root node, all but one of the nodes $(x,y)$,
for an appropriately chosen $y$, in class $S_x$, and the nodes
$(\alpha^i,\alpha^i)$, for $i=0,1,2,\dots,p^s-2$.  We have chosen
$2p^s$ active variable nodes in all.
 
The variable nodes $(1,1),
(\alpha,\alpha),\dots,(\alpha^{p^s-2},\alpha^{p^s-2})$ are
connected to constraint nodes \[(\alpha^k,0)'_c, (\alpha^k,1+\alpha^k)'_c,
(\alpha^k,\alpha+\alpha^{k+1})'_c,\dots, (\alpha^k,\alpha^j+\alpha^{k+j})'_c,\dots,(\alpha^{k},\alpha^i+\alpha^{k+p^s-2})'_c,\]
respectively, in class $S_{\alpha^k}'$ of constraint nodes in
layer $L_3$. These nodes are either all distinct or all equal to
$(\alpha^k,0)'_c$ since $\alpha^r+\alpha^{k+r}=
\alpha^t+\alpha^{k+t}$ if and only if either $r=t$ or
$1+\alpha^k=0$. Since $1+\alpha^k$ is zero for exactly one value of
$k\in \{0,1,2,\dots,p^s-2\}$, we have that the variable nodes
$(\alpha^i,\alpha^i)$, for $i=0,1,2,\dots,p^s-2$, are connected to
distinct constraint nodes in all but one class $S_{\alpha^{k*}}$
and that, in $S_{\alpha^{k*}}$, they are all connected to the constraint node $(\alpha^{k*},0)'_c$.
(Note that $k*$ satisfies $1+\alpha^{k*}=0$.) We let $y=\alpha^{k*}$. Therefore, the set of active
variable nodes includes all nodes of the form $(x,t)$, for
$t=0,1,\alpha,\alpha^2,\dots$, excluding node $(x,y)$.

Since the code is $p$-ary, assign the following values to the
chosen set of active variable nodes: assign the value 1 to the
root variable node and to all the active variable nodes in class
$S_x$, and assign the value $p-1$ to the active variable nodes
$(\alpha^i,\alpha^i)$, for $i=0,1,2,\dots,p^s-2$. It is now easy
to verify that all the constraints are satisfied. Thus,
$d_{\min}\le 2p^s$. From Theorem~\ref{type2ell3-girth-thm} and Lemmas~\ref{q_ary_dmin} and
\ref{q_ary_wmin}, we have $T(p^s+1,6)\le w_{\min}\le d_{\min}\le 2p^s$.\\
\end{proof}
\vspace{0.1in}

For degrees $d=p^s+1$, $p>2$, treating the Type II $\ell=3$ LDPC
constraint graphs as binary LDPC codes, yields $[n,1,n]$ repetition codes, where
$n=p^{2s}+p^s+1$, $d_{\min}=n$, and dimension is $1$. However,
when the Type II $\ell=3$ LDPC constraint graphs, for degrees
$d=p^s+1$, $p>2$, are treated as $p$-ary LDPC codes, we believe
that the distance $d_{\min}\ge p^s+3$, and that this
bound is in fact tight. We also suspect
that the minimum pseudocodeword weights (on the $p$-ary symmetric
channel) are much closer to the minimum distances for these $p$-ary LDPC codes.
\\
\vspace{0.1in}

\subsection{$\ell = 4$}
Figure~\ref{type2g8d3_graph} provides an example of a Type II $\ell=4$ LDPC constraint
graph with degree $d=2+1=3$ and girth $g=8$.
For $d=p^s+1$, $p$ a prime and $s$ a positive integer, a $d$
regular tree $T$ is enumerated from a root (variable) node for
$\ell=4$ layers $L_0,L_1,L_2,L_3$.
\begin{enumerate}
\item The  nodes
  in $L_0, L_1,$ and $L_2$ labeled as in the $\ell = 3$ case. The
  constraint nodes in $L_3$ are labeled as follows: The
  constraint nodes  descending from variable node $(x,j)$, for $j
  =0,1,,\alpha,\dots,\alpha^{p^s-2}$, are labeled
  $(x,j,0)_c,(x,j,1)_c,\dots, (x,j,\alpha^{p^s-2})_c$, the constraint
  nodes descending from variable node $(i,j)$, for $i,j=0,1,,\alpha,\dots,\alpha^{p^s-2}$, are labeled  $(i,j,0)_c,(i,j,1)_c,\dots, (i,j,\alpha^{p^s-2})_c$.

\item A final layer $L_{\ell}=L_4$ of $(d-1)^{\ell-1}=p^{3s}$
  variable nodes is introduced. The $p^{3s}$ variable nodes in $L_4$ are labeled
  as  $(0,0,0)', (0,0,1)',\dots, (0,0,\alpha^{p^s-2})'$, $(0,1,0)',(0,1,1)',\dots,
  (0,1,\alpha^{p^s-2})'$, $\dots$,
  $(\alpha^{p^s-2},0,0)',$ $(\alpha^{p^s-2},0,1)',$ $\dots,$ $(\alpha^{p^s-2},0,\alpha^{p^s-2})'$, $\dots,$   $(\alpha^{p^s-2},\alpha^{p^s-2},0)'$, $(\alpha^{p^s-2},\alpha^{p^s-2},1)',$ $\dots,$ $(\alpha^{p^s-2},\alpha^{p^s-2},\alpha^{p^s-2})'$.
(Note that the `$'$' in the labeling refers to nodes that are not
in the tree $T$ and the subscript `$c$' refers to constraint nodes.)
\item For $0\le i\le p^s-1$, $0\le j\le p^s-1$, connect the constraint
  node $(x,i,j)_c$ to the variable nodes \[(i,j,0)', (i,j,1)',(i,j,\alpha)',\dots,(i,j,\alpha^{p^s-2})'.\]
 
\item To connect the remaining constraint nodes in $L_3$ to the
  variable nodes in $L_4$, we first define a function $f$. For $i,j,k,t=0,1,\alpha,\dots,\alpha^{p^s-2}$  let
    \[f:\mathbb{F}\times\mathbb{F}\times
    \mathbb{F}\times\mathbb{F}\rightarrow \mathbb{F}\]
\[(i,j,k,t)\mapsto y,\]
  be an appropriately chosen function, that we will define later for
    some specific cases of the Type II $\ell=4$ construction.
Then,   for $i,j,k=0,1,\alpha,\dots,\alpha^{p^s-2}$ , connect
the constraint node $(i,j,k)_c$ in $L_3$ to the following variable
nodes in $L_4$
\[\hspace{-0.3in}(0,k+i\cdot 0,f(i,j,k,0))',\ (1,k+i\cdot 1,f(i,j,k,1))',\
(\alpha,k+i\cdot \alpha,f(i,j,k,\alpha))',\ ..,\ (\alpha^{p^s-2},k+i\cdot \alpha^{p^s-2},f(i,j,k,\alpha^{p^s-2}))'.\] 
(Observe that the second index corresponds to the linear map
defined by the array $M_{i}$ defined in Section 2.B. Further, note that if $f(i,j,k,t)=j+ i\cdot t$, then 
the resulting graphs obtained from the above set of
connections have girth at least six. However, there are other functions $f(i,j,k,t)$ for
    which the resulting graphs have girth exactly eight, which is
    the best possible when $\ell=4$ in this construction. At this
    point, we do not have a closed form expression for the
    function $f$ and we only provide details for specific
    cases below. (These cases were verified using the MAGMA
    software \cite{maxxu}.)\\
\end{enumerate}

The  Type II, $\ell = 4$, LDPC codes have girth eight, minimum distance $d_{\min}\ge 2(p^s+1)$, and blocklength $N=1+p^s+p^{2s}+p^{3s}$.
(We believe that the tree bound on the minimum distance is  met
for most of the Type II, $\ell=4$, codes, i.e. $d_{\min}=w_{\min}=2(p^s+1)$.)
For $d=3$, the Type II, $\ell=4$, LDPC constraint graph as
shown in Figure~\ref{type2g8d3_graph} corresponds to the
$(2,2)$-Finite-Generalized-Quadrangles-based LDPC (FGQ LDPC) code of
\cite{vo01p}; the function $f$ used in constructing this
example is defined by $f(i,j,k,t)=j+(i+1)\cdot t$, i.e., the
map defined by the array $M_{i+1}$. The orthogonal arrays used for constructing this code are

\[\tiny 
\hspace{-0in} M_{0}=\left[\begin{array}{cc}
    0&0\\
    1&1
\end{array}\right], \ M_{1}=\left[\begin{array}{cc}
0&1\\
1&0
\end{array}\right].\]

We now state some results concerning the choice of the
function $f$.

\begin{enumerate}
\item The Type II $\ell=4$ construction results in incidence graphs of
finite generalized quadrangles for appropriately chosen functions
$f$. These graphs have girth $8$ and diameter $4$.

\item For some specific cases, examples of the function $f(i,j,k,t)$ that
  resulted in a girth $8$ graph is  given in
  Table~\ref{f_func_type2ell4}. (Note that for the second entry
  in the table, the function $g:GF(4)\rightarrow GF(4)$
  is defined by the following maps: $0\mapsto 1$, $1\mapsto
  \alpha$, $\alpha\mapsto \alpha^2$, and $\alpha^2\mapsto 0$.) We
  have not been able to find a general relation or a closed form
  expression for $f$ yet. 
{\scriptsize
\begin{table}[h]
\begin{center}
\begin{tabular}{|c|c|c|c|c|}
\hline
$p$&$s$&elements&degree&$f(i,j,k,t)$\\
& & of $GF(p^s)$&$d=p^s+1$&\\
\hline \hline
2&1&$\{0,1\}$&3&$j+(i+1)\cdot t$\\
2&2&$\{0,1,\alpha,\alpha^2\}$&5&$j+g(i)\cdot t$\\
3&1&$\{0,1,2\}$&4&$i\cdot(k+2\cdot i\cdot t)+j$\\
3&2&$\{0,1,\alpha,..,\alpha^{7}\}$&10&$i\cdot(k+\alpha\cdot i\cdot t)+j$\\
5&1&$\{0,1,2,3,4\}$&6&$i\cdot(k+3\cdot i\cdot t)+j$\\
7&1&$\{0,1,2,..,6\}$&8&$i\cdot(k+4\cdot i\cdot t)+j$\\
\hline
\end{tabular}
\caption{The function $f$ for the Type II $\ell=4$  construction.}
\label{f_func_type2ell4}
\end{center}
\end{table}
}

\item  For the above set of functions, the resulting Type II $\ell=4$
   LDPC constraint graphs have minimum distance meeting the tree
   bound, when $p=2$, i.e., $d_{\min}=w_{\min}=2(2^s+1)$. We
   conjecture that, in general, for degrees $d=2^s+1$, the Type II $\ell=4$
   girth eight LDPC constraint graphs have $d_{\min}=w_{\min}=T(2^s+1,8)=2(2^s+1)$.
\item For degrees $d=p^s+1$, $p>2$, we expect the corresponding $p$-ary
  LDPC codes from this construction to have minimum distances
  $d_{\min}$ either equal or very close to the tree bound. Hence,
  we also expect 
  the corresponding minimum pseudocodeword weight $w_{\min}$ to be close to $d_{\min}$.

\end{enumerate}

The above results were verified using MAGMA and computer simulations.\\

\subsection{Remarks}
It is well known in the literature that finite generalized
polygons (or, $N$-gons) of order $p^s$ exist
\cite{ma98b2}. A finite generalized $N$-gon is a non-empty
point-line geometry, and consists of a
set $\mathcal{P}$ of points and a set $\mathcal{L}$ of lines such
that the incidence graph of this geometry is a bipartite graph of
diameter $N$ and girth $2N$. Moreover, when each point is incident on
$t+1$ lines and each line contains $t+1$ points, the order of the
$N$-gon is said be to $t$. The Type II $\ell=3$ and $\ell=4$
constructions yield finite generalized $3$-gons and $4$-gons,
respectively, of order $p^s$. These are essentially finite projective planes and
finite generalized quadrangles. The Type II construction can be similarly extended to larger
$\ell$. We believe that finding the right connections for
connecting the nodes
between the last layer in $T$ and the final layer will yield
incidence graphs of these other finite generalized
polygons. For instance, for $\ell=6$ and $\ell=8$, the
construction can yield finite generalized hexagons and
finite generalized octagons, respectively. We conjecture that the
incidence graphs of generalized $N$-gons  yield 
LDPC codes with  minimum pseudocodeword
weight $w_{\min}$ very close to the corresponding minimum
distance $d_{\min}$ and particularly, for generalized $N$-gons of order $2^s$,
the LDPC codes have $d_{\min}=w_{\min}=T(2^s+1,2N)$.\\ \vspace{0.1in}

\section{Simulation Results}
\subsection{Performance with min-sum iterative decoding}
Figures \ref{type1_perf}, \ref{type1B_perf}, \ref{type2g6_perf},
\ref{type2g8_perf} show the bit-error-rate performance of Type
I-A, Type I-B, Type II $\ell=3$ (girth six), and Type II $\ell=4$
(girth eight) LDPC
codes, respectively, over the binary input additive white Gaussian
noise channel (BIAWGNC) with min-sum (not, sum-product!)
iterative decoding, as a function of the channel signal to noise
ratio (SNR) $E_b/N_o$. The performance of regular or semi-regular randomly constructed LDPC codes of comparable rates and block lengths are also shown. (All of the random LDPC codes compared in this paper have a variable node degree of three and are constructed from the online LDPC software available at {\tt \scriptsize http://www.cs.toronto.edu/$\tilde{}$ radford/ldpc.software.html}.) 

Figure \ref{type1_perf} shows that the Type I-A LDPC codes
perform substantially better than their random counterparts for
all the codes shown. In
these simulations, a maximum of 10000 iterations of min-sum
decoding were allowed. 

Figure \ref{type1B_perf} reveals that the Type I-B LDPC codes
perform better than comparable random LDPC codes at short
block lengths; but as the block lengths increase, the random LDPC
codes tend to perform better in the waterfall region. Eventually
however, as the SNR increases, the Type I-B LDPC codes outperform
the random ones  and, unlike the random codes, they do not have a
prominent error floor. For short block lengths (below 4000), a
maximum of 200 decoding iterations were allowed whereas for the
longer block lengths, only up to 20 decoding iterations were allowed.
The block length 16385, rate 0.951, Type I-B LDPC code performs significantly
better than the corresponding random LDPC code of the same block
length and rate. 

Figure \ref{type2g6_perf} reveals that the
performance of Type II $\ell=3$ (girth-six) LDPC codes is also
significantly better than comparable random codes; these codes
correspond to the two dimensional PG-LDPC codes of
\cite{li05}. The Type II codes significantly outperform the corresponding
random LDPC codes of comparable parameters at short block lengths. At longer block lengths,
the random codes tend to perform better in the waterfall region
due to their superior minimum distances.
A maximum of 200 iterations were allowed for short
block length codes (block lengths below 4000) whereas only 20
iterations were allowed for the block length 4161 code. Here
again, the Type II codes do not reveal a prominent error as the
corresponding random LDPC codes do, and at longer block lengths, they outperform the
corresponding random LDPC codes in the high SNR regime.

Figure \ref{type2g8_perf} indicates the performance of
Type II $\ell=4$ (girth-eight) LDPC codes; these codes perform
comparably to random codes at short block lengths, but at large
block lengths, the random codes perform better as they have larger
relative minimum distances compared to the Type II $\ell=4$
(girth-eight) LDPC codes. In these simulations, a maximum of 200
decoding iterations were allowed for all the codes shown. The
performance of the Type II $\ell=4$ LDPC codes reveal a similar
trend as that of the  Type II $\ell=3$ LDPC codes in
Figure~\ref{type2g6_perf}.

As a general observation, min-sum iterative decoding of most of
the tree-based LDPC codes (particularly, Type I-A, Type II,  and
some Type I-B) presented here did not typically reveal detected
errors, i.e., errors caused due to the decoder failing to
converge to any valid codeword within the maximum specified
number of iterations, which was set to 200 for short block length
codes (and 20 for longer blocklength codes) in these simulations. Detected errors are caused primarily due to
the presence of pseudocodewords, especially those of minimum
weight. We think that the relatively low occurrences of detected errors with iterative
decoding of these LDPC codes is primarily due to their
good\footnote{i.e., relative to the minimum distance $d_{\min}$.}
minimum pseudocodeword weight $w_{\min}$. \\

\subsection{Performance of Type I-B and Type II LDPC codes with
  sum-product iterative decoding}
Figures~ \ref{type1B_SPperf},  \ref{type2l3_SPperf}, and
\ref{type2l4_SPperf} show the performance of the Type I-B, Type
II $\ell=3$ and Type II $\ell=4$, respectively, LDPC codes with
sum-product iterative decoding for the BIAWGNC. The performance is shown only for a few codes from each
construction. The main observation from these
performance curves is that the the tree-based LDPC codes perform
relatively much better than random LDPC codes of comparable
parameters when the decoding is sum-product instead of the min-sum algorithm.
 Although the Type I-B LDPC codes perform a little inferior to
 their random counterparts in the waterfall region when the
 block length is large, the gap
 between the performances of the random and the Type I-B LDPC
 codes is much smaller with sum-product decoding than
 with min-sum decoding. (Compare Figures~\ref{type1B_perf}
 and \ref{type1B_SPperf}.) Once again, a maximum of 200 decoding
 iterations were performed for block lengths below 4000 and 20
 decoding iterations were performed for the block length 4097 LDPC
 code.

Similarly, comparing Figures
 ~\ref{type2l3_SPperf} and \ref{type2g6_perf}, we see that the
 Type II $\ell=3$  LDPC codes perform relatively much better than their random
 counterparts with sum-product decoding than with min-sum
 decoding. They outperform the corresponding random
 LDPC codes at block lengths below 1000, whereas at the longer
 block lengths, the random codes perform better than the Type II
 codes in the waterfall region. Figure~\ref{type2l4_SPperf}, in comparison with
 Figure~\ref{type2g8_perf}, shows a similar trend in performance
 of Type II $\ell=4$ (girth 8) LDPC codes with sum-product
 iterative decoding.

Note that the simulation results for the min-sum and sum-product
decoding correspond to the case when the LDPC codes resulting
from constructions Type I and Type II were treated as binary LDPC
codes for all choices of degree $d=p^s$ or $d=p^s+1$. We will now
examine the performance when the codes are treated as $p$-ary codes if
the corresponding degree in the LDPC constraint graph is $d=p^s$
(for Type I-B)
or $d=p^s+1$ (for Type II). (Note that this will affect only the performances of those codes for which $p$ is not equal to two.) \\

\subsection{Performance of $p$-ary Type I-B and Type II LDPC
  codes over the $p$-ary symmetric channel}
We examine the performance of the $p$-ary LDPC codes obtained from
the Type I-B and Type II constructions on the
$p$-ary symmetric channel instead of the AWGN channel. The
$p$-ary symmetric channel is shown in Figure~\ref{q_ary_sym}. An error occurs with probability $\epsilon$, the channel
transition probability. Figures~\ref{type1B_pary_perf},
\ref{type2l3_pary_perf}, and \ref{type2l4_pary_perf} show the
performance of Type I-B, Type II $\ell=3$ and Type II $\ell=4$,
$3$-ary LDPC codes, respectively, on the $3$-ary symmetric
channel with sum-product iterative decoding. A maximum of 200
sum-product decoding iterations were performed. The parity check
matrices resulting from the the Type I-B and Type II constructions are
considered to be matrices over the field $GF(3)$ and sum-product
iterative decoding is implemented as outlined in
\cite{da99t}. The corresponding plots show the information
symbol error rate as a function of the channel transition
probability $\epsilon$. In Figure~\ref{type1B_pary_perf}, the
performance of $3$-ary Type I-B LDPC codes obtained for degrees $d=3$, $d=3^2$,
and $d=3^3$,  is shown and compared with the performance of random
$3$-ary LDPC codes of comparable rates and block lengths. (To
make a fair comparison, the random LDPC codes also have only zeros
and ones as entries in their parity check matrices. It has been
observed in \cite{da99t} that choosing the
non-zero entries in the parity check matrices of non-binary codes
cleverly can yield some
performance gain, but this avenue was not explored in these
simulations.) In Figure~\ref{type2l3_pary_perf}, the performance
of $3$-ary Type II $\ell=3$ (girth six) LDPC codes obtained for
degrees $d=3+1$, $d=3^2+1$, and $d=3^3+1$, is shown and compared
with random $3$-ary LDPC codes.  Figure~\ref{type2l4_pary_perf}
shows the analogous performance of $3$-ary Type II $\ell=4$
(girth eight) LDPC codes obtained for degrees $d=3+1$ and
$d=3^2+1$. In all these plots, it is seen that the tree-based
constructions perform comparably or better than
random LDPC codes of similar rates and block lengths. (In some
cases, the performance of the tree-based constructions is
significantly better than that of random LDPC codes (example,
Figure~\ref{type2l4_pary_perf}).)  

The simulation results show that the tree-based constructions
yield LDPC codes with a wide range of rates and block lengths that perform very well with
iterative decoding.\\
\vspace{0.1in}

\section{Conclusions}

The Type I construction yields a family of LDPC codes that, to the best of 
our knowledge, do not correspond to any of the LDPC codes obtained from 
finite geometries or other geometrical objects. It would be interesting to 
extend the Type II construction to more layers as described at the end of 
Section 5, and to extend the Type IA construction by relaxing the girth 
condition. In addition, these codes may be amenable to efficient 
tree-based encoding procedures. A definition for the pseudocodeword weight 
of $p$-ary LDPC codes on the $p$-ary symmetric channel was also derived, 
and an extension of the tree bound in \cite{ke05u} was obtained. This led 
to a useful interpretation of the tree-based codes, including the 
projective geometry LDPC codes, for $p > 2$. The tree-based constructions 
presented in this paper yield a wide range of codes that perform well when 
decoded iteratively, largely due to the maximized minimum pseudocodeword 
weight. While the tree-based constructions are based on pseudocodewords 
that arise from the graph-cover's polytope of \cite{ko03p} and aim to 
maximize the minimum pseudocodeword weight of pseudocodewords in this set, 
they do not consider all pseudocodewords arising on the min-sum iterative 
decoder's computation tree \cite{wi96t}. Nonetheless, having a large 
minimal pseudocodeword weight in this set necessarily brings the 
performance of min-sum iterative decoding of the tree-based codes closer 
to the maximum-likelihood performance. However, it would be interesting to 
find other design criteria that account for pseudocodewords arising on the 
decoder's computation tree.

Furthermore, since the tree-based constructions have the minimum 
pseudocodeword weight and the minimum distance close to the tree bound, 
the overall minimum distance of these codes is relatively small. While 
this is a first step in constructing LDPC codes having the minimum 
pseudocodeword weight $w_{\min}$ equal/almost equal to the minimum 
distance $d_{\min}$, constructing codes with larger minimum distance, 
while still maintaining $d_{\min} = w_{\min}$, remains a challenging 
problem. \\ \vspace{0.1in}

\appendix

\subsection{\sc Pseudocodeword weight for $p$-ary LDPC codes on
    the $p$-ary symmetric channel}

Suppose the all-zero codeword is sent across a $p$-ary symmetric
channel and the vector ${\bf r}=(r_0,r_1,\dots,r_{n-1})$ is
received. Then errors occur in positions where $r_i\ne 0$. Let
$S=\{i |\ r_i\ne 0\}$ and let $S^c=\{i |\ r_i=0\}$.
The distance between ${\bf r}$ and a pseudocodeword $F$ is
defined as
\begin{equation}
d({\bf r},F)=\sum_{i=0}^{n-1}\sum_{k=0}^{p-1}\chi(r_i\ne
k) f_{i,k}, \label{dist}\end{equation}
where $\chi(P)$ is an indicator function that is equal to $1$
if the proposition $P$ is true and is equal to $0$ otherwise.

The distance between ${\bf r}$ and the all-zero codeword ${\bf
  0}$ is \[d({\bf r},{\bf 0})=\sum_{i=0}^{n-1}\chi(r_i\ne 0) \]
which is the Hamming weight of ${\bf r}$ and can be obtained from
equation (\ref{dist}).

The iterative decoder chooses in favor of $F$ instead of the
all-zero codeword ${\bf 0}$ when $d({\bf r},F)\le d({\bf r},{\bf
  0})$.
 That is, if 
\[ \sum_{i\in S^c}(1-f_{i,0}) +\sum_{i\in S}(1-f_{i,r_i})\le
\sum_{i\in S}1\]
The condition for choosing $F$ over the all-zero codeword reduces
to 
\[ \Big{\{} \sum_{i\in S^c}(1-f_{i,0})\le \sum_{i\in
  S}f_{i,r_i}\Big{\}}\]
 Hence, we define the weight of a pseudocodeword $F$ in the
 following manner.

 Let $e$ be a number such that the sum of the $e$ largest components in
the matrix $F'$, say,
$f_{i_1,j_1},f_{i_2,j_2},\dots,f_{i_e,j_e}$, exceeds $\sum_{i\ne i_1,i_2,\dots,i_e}(1-f_{i,0})$.
Then the weight of $F$ on the $p$-ary symmetric channel is
defined as 
\[ w_{PSC}(F) =\left\{ \begin{array}{cc}
           2e, & \mbox{if } f_{i_1,j_1}+\dots +f_{i_e,j_e} =\sum_{i\ne i_1,i_2,\dots,i_e}(1-f_{i,0})\\
2e-1, & \mbox{if } f_{i_1,j_1}+\dots +f_{i_e,j_e} > \sum_{i\ne i_1,i_2,\dots,i_e}(1-f_{i,0})
\end{array} \right  . 
\] 
Note that in the above definition, none of the $j_k$'s, for
$k=1,2,\dots,e$, are equal to zero, and all the $i_k$'s, for
$k=1,2,\dots,e$, are distinct. That is, we choose at most one
component in every row of $F'$ when picking the $e$
largest components. The received vector ${\bf r}=(r_0,r_1,\dots,r_{n-1})$ that has
the following components: 
$r_{i_1}=j_1, r_{i_2}=j_2, \dots, r_{i_e}=j_e$, $r_i=0$, for
$i\notin \{i_1,i_2,\dots,i_e\}$, will cause the decoder to make
an error and  choose
$F$ over the all-zero codeword.\\

\subsection{\sc Proof of Lemma~\ref{q_ary_wmin}}

\begin{proof} 
{\begin{center}
\begin{figure}[h]
\centering{
            \begin{minipage}[b]{0.35\linewidth} 
            \centering
                 {
             \resizebox{1in}{1in}{\vspace{1in}\includegraphics{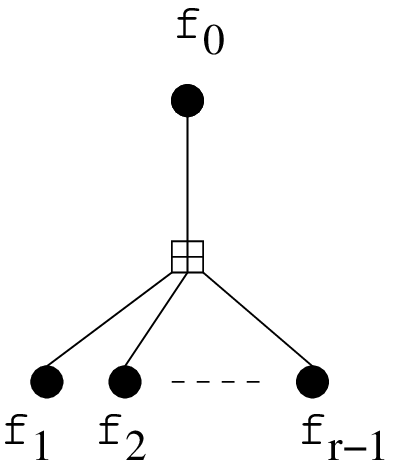}\vspace{0in}}
             \caption{Single constraint code.}\label{spc}
\vspace{0.2in}
             $({1-f_{i,0}})\le \sum_{j\ne i}(1-f_{j,0})$\vspace{0.1in}}
         \end{minipage}
            \hspace{0.1in}
      \begin{minipage}[b]{0.6\linewidth}\vspace{0in}
    \centering{
            \resizebox{1.8in}{1.8in}{\includegraphics{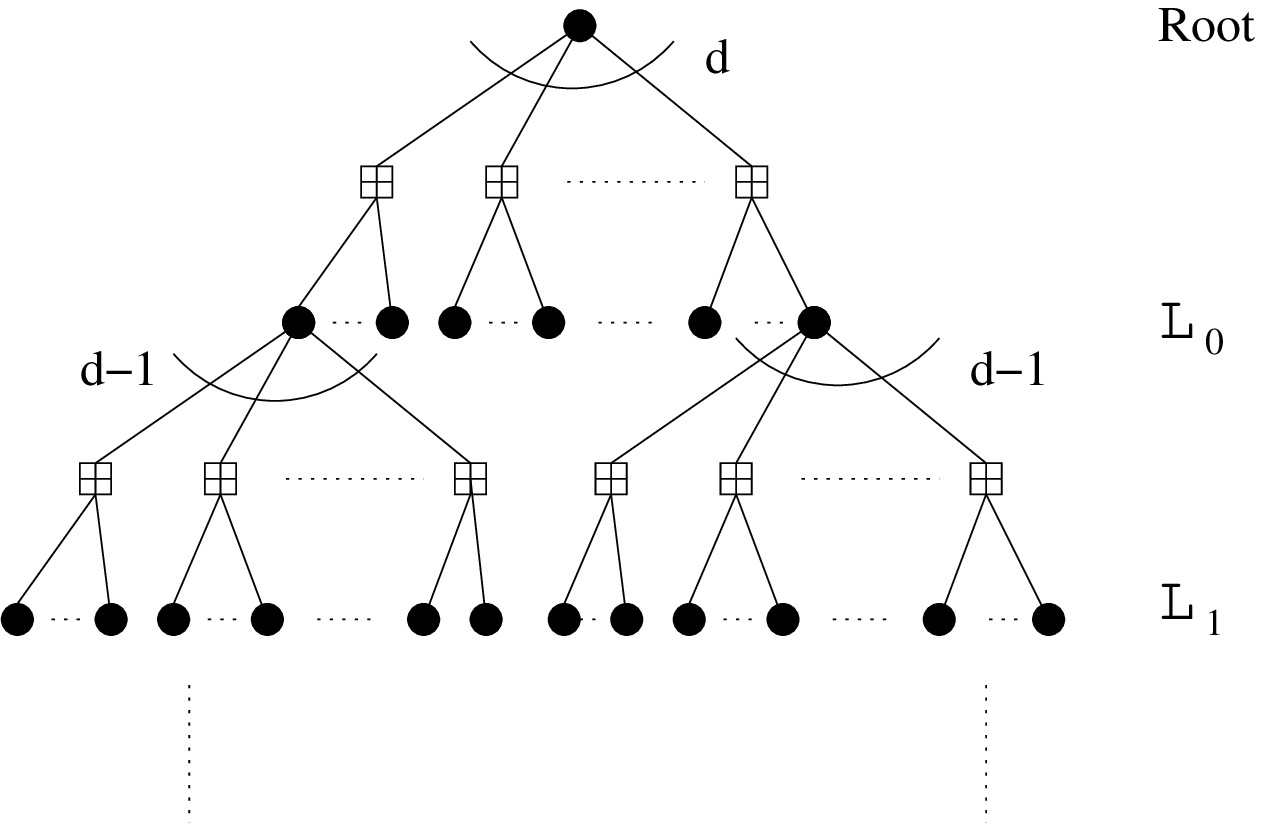}}}
            \caption{Local tree structure for a $d$-left regular graph.}
        \label{tree}
$d (1-f_{0,0}) \le \sum_{j\in
          L_0}(1-f_{j,0})$,\vspace{-0.05in}\\ 
 $d(d-1)(1-f_{0,0})\le \sum_{j\in L_1}(1-f_{j,0})$\vspace{-0.05in}\\
   $:$\vspace{-0.05in}\\
           \end{minipage}}
\end{figure}
\end{center}
}

\underline{Case:} $\frac{g}{2}$ odd. Consider a single constraint node
with $r$ variable node neighbors as shown in
Figure~\ref{spc}. Then, for $i=0,1,\dots, r-1$ and $k=0,1,\dots,p-1$,
the following inequality holds: 
\[ f_{i,k} \le \sum_{j\ne i} \sum_{\sigma_j: \sum \sigma_j+k=0 \mbox{ mod } p} \frac{
  f_{j,\sigma_j}\sigma_j}{\sum \sigma_j} \]
where the middle summation is over all possible assignments
$\sigma_j\in \{ 0,1,\dots,p-1\}$ to the variable nodes $j\ne i$ such
that $k+\sum_{j\ne i} \sigma_j =0 \mbox{ mod } p$, i.e., this is a
valid assignment for the constraint node. The innermost summation
in the denominator is  over all $j\ne i$.

However, for $i=0,1,\dots, r-1$, the following (weaker) inequality
also holds:
\begin{eqnarray}
(1-f_{i,0})\le   \sum_{j\ne i} (1-f_{j,0})
\label{eqn_spc}
\end{eqnarray}

Now let us consider a $d$-left regular LDPC constraint graph
representing a $p$-ary LDPC code. We will enumerate the LDPC
constraint graph as a tree from an arbitrary root variable node,
as shown in Figure~\ref{tree}. Let $F$ be a pseudocodeword matrix
for this graph. Without loss of generality, let us
assume that the component $(1-f_{0,0})$ corresponding to the root
node is the maximum among all $(1-f_{i,0})$ over all $i$.

Applying the inequality in (\ref{eqn_spc}) at every constraint
node in first constraint node layer of the tree, we obtain

\[ d(1-f_{0,0}) \le \sum_{j\in L_0}(1-f_{j,0}),\] where $L_0$
corresponds to variable nodes in first level of the
tree. Subsequent application of the inequality in (\ref{eqn_spc})
to the second layer of constraint nodes in the tree yields
 \[  d(d-1)(1-f_{0,0})\le \sum_{j\in L_1}(1-f_{j,0}),\]
 Continuing this process  until layer $L_{\frac{g-6}{4}}$,
we obtain 
\[d(d-1)^{\frac{g-6}{4}}(1-f_{0,0})\le \sum_{j\in  L_{\frac{g-6}{4}}}(1-f_{j,0})\]

Since the LDPC graph has girth $g$, the variable nodes up to level
$L_{\frac{g-6}{4}}$ are all distinct. The above
inequalities yield:
    \begin{equation}
[1+d+d(d-1)+\dots+d(d-1)^{\frac{g-6}{4}}](1-f_{0,0}) \le \sum_{i\in
\{0\}\cup L_0\cup\dots L_{\frac{g-6}{4}}} (1-f_{i,0})\le
\sum_{\mbox{ all }i}(1-f_{i,0})
\label{fin_ineq}
\end{equation} 

Let $e$ the smallest number such that there are $e$ maximal
components  $f_{i_1,j_1}$, $f_{i_2,j_2}, f_{i_3,j_3},\dots,
f_{i_e,j_e}$, for $i_1,i_2,\dots,i_e$ all distinct and
$j_1,j_2,\dots,j_e \in \{1,2,\dots,p-1\}$,  in $F'$ (the sub-matrix of $F$
excluding the first column in $F$) such that 
\[f_{i_1,j_1}+f_{i_2,j_2}+\dots+f_{i_e,j_e}\ge \sum_{i\notin
  \{i_1,i_2,i_3,\dots,i_e\}}(1-f_{i,0}) \]

Then, since none of the $j_k$'s, $k=1,2,\dots,e$, are zero, we
clearly have
\[ (1-f_{i_1,0})+(1-f_{i_2,0})+\dots+(1-f_{i_e,0})\ge f_{i_1,j_1}+f_{i_2,j_2}+\dots+f_{i_e,j_e}
\ge \sum_{i\notin
  \{i_1,i_2,i_3,\dots,i_e\}}(1-f_{i,0})\]
Hence we have that
\[2((1-f_{i_1,0})+(1-f_{i_2,0})+\dots+(1-f_{i_e,0})) \ge \sum_{\mbox{all }
  i}(1-f_{i,0})\]

We can then lower bound this further using the inequality in
(\ref{fin_ineq}) as 
\[2((1-f_{i_1,0})+(1-f_{i_2,0})+\dots+(1-f_{i_e,0}))\ge
[1+d+d(d-1)+\dots+d(d-1)^{\frac{g-6}{4}}](1-f_{0,0})\]
Since we assumed that $(1-f_{0,0})$ is the maximum among 
$(1-f_{i,0})$ over all $i$,   we have
\[2e(1-f_{0,0})\ge 2((1-f_{i_1,0})+(1-f_{i_2,0})+\dots+(1-f_{i_e,0}))\ge
[1+d+d(d-1)+\dots+d(d-1)^{\frac{g-6}{4}}](1-f_{0,0})\]
This yields the desired bound \[w_{PSC}(F)= 2e\ge 1+d+d(d-1)+\dots+d(d-1)^{\frac{g-6}{4}}\]
Since the pseudocodeword $F$ was arbitrary, we also have
$w_{min}\ge 1+d+d(d-1)+\dots+d(d-1)^{\frac{g-6}{4}}$. The case
$\frac{g}{2}$ even is treated similarly. \\
\end{proof}
\vspace{0.1in}

\subsection{\sc Table of code parameters}
The  code parameters resulting from the tree-based constructions
are summarized in
Tables~\ref{table_typeIa},~\ref{table_typeIb},~\ref{table_typeII3},
and ~\ref{table_typeII4}. Note that $^*$ indicates an upper bound instead of the exact
minimum distance (or minimum pseudocodeword weight) since it was
computationally hard to find the distance (or pseudoweight) for
those cases. Similarly, for cases where it was computationally
hard to get any reasonable bound the minimum pseudocodeword
weight, the corresponding entry in the table is left empty. The
lower bound on $w_{\min}$ seen in the tables corresponds to the
tree bound (Theorem~\ref{thm1}).  It is observed that when the
codes resulting from the construction are treated as $p$-ary
codes rather than binary codes when the corresponding degree in
the LDPC graph is $d=p^s$ (for Type I-B)
or $d=p^s+1$ (for Type II), the resulting rates obtained are much superior; we also believe
that the minimum pseudocodeword weights  (on the $p$-ary
symmetric channel) are much closer to the minimum distances for
these $p$-ary LDPC codes.\\
\vspace{0.1in}

\def\cprime{$'$} \def\cprime{$'$}


{\tiny
\begin{table}
\begin{center}
\begin{tabular}{|c|c|c|c|c|c|c|c|c|c|}
\hline
No. of layers in $T$&block length&degree&dimension&rate&$d_{\min}$&$w_{\min}$&tree&girth&diameter\\
$\ell$&$n$&$d=3$& & & & &lower-bound&$g$&$\delta$\\
\hline \hline
3&10&3&4&0.4000&4&4&4&6&5\\
4&22&3&4&0.1818&8&$8^*$&6&8&7\\
5&46&3&10&0.2173&10&10&10&10&9\\
6&94&3&14&0.1489&20&$20^*$&18&12&11 \\
{\bf 7}&{\bf 190}&3&25 & 0.1315&24 &$\ge 18$&18&{\bf 12}&13 \\
\hline
\end{tabular}
\caption{Summary of Type I-A code parameters.}
\label{table_typeIa}
\end{center}
\vspace{-0.35in}
\begin{center}
\begin{tabular}{|c|c|c|c|c|c|c|c|c|c|c|c|}
\hline
$p$&$s$&block length&degree&dimension&rate&$d_{\min}$&$w_{\min}$&tree&girth&diameter&code\\
&   &$n=p^{2s}+1$&$d=p^s$& & & & &lower-bound&$g$&$\delta$&alphabet\\
\hline \hline
2&1&5&2&1 &0.2000 &5 &5 &4&8&4&binary \\ \hline
3&1&10&3&3&.3000&4&4&4&6& 5&binary\\
 & & & &(2)&(0.2000)&(6)&( )& & & &3-ary\\ \hline
2&2&17&4&5&0.2941&6&$6^*$&5&6& 5&binary\\ \hline
5&1&26&5&7&0.2692&8&$8^*$&6&6& 5&binary\\ 
 & & & &(7)&(0.2692)&(10)&( )& & & &5-ary\\ \hline
7&1&50&7&11&0.2200&12&$12^*$&8&6&5 &binary\\
 & & & &(16)&(0.3200)&($15$)&( )& & & &7-ary\\ \hline
2&3&65&8&31&0.4769&$10$&$10^*$&9&6&5 &binary\\ \hline
3&2&82&9&15&0.1829&16&$\ge 10$&10&6&5 &binary\\
 & &  & &(38)&(0.4634)&$(15)$& ( )& & & &3-ary\\ \hline
11&1&122&11&19&0.1557&$20^*$&$\ge 12$&12&6&5 &binary\\
 & & & &(46)&(0.3770)&($30^*$)&( )& & & &11-ary\\ \hline
2&4&257&16&161&0.6264&$20^*$&$\ge 17$&17&6&5 &binary \\ \hline
5&2&626&25&47&0.075&$48^*$& $\ge 26$ &26&6&5 &binary\\
 & & & &(377)&(0.6022)&($90^*$)&( )& & & &5-ary\\ \hline
3&3&730&27&51&0.0698&$52^*$& $\ge 28$ &28&6&5 &binary\\
 & & & &(488)&(0.6684)&($60^*$)&( )& & & &3-ary\\ \hline
2&5&1025&32&751&0.7326&$40^*$&$\ge 33$&33&6&5 &binary\\ \hline
7&2&2404&49&95&0.0395&$96^*$& $\ge 50$&50&6&5 &binary\\
 & & & &(1572)&(0.6536)&($216^*$)&( )& & & &7-ary\\
\hline
\end{tabular}
\caption{Summary of Type I-B code parameters.}
\label{table_typeIb}
\end{center}
\end{table}
}

{\tiny
\begin{table}
\begin{center}
\begin{tabular}{|c|c|c|c|c|c|c|c|c|c|c|c|}
\hline
$p$&$s$&block length&degree&dimension&rate&$d_{\min}$&$w_{\min}$&tree&girth&diameter&code\\
 &   &$n=1+p^s+p^{2s}$&$d=p^s+1$& & & & &lower-bound&$g$&$\delta$&alphabet\\
\hline \hline
2&1&7&3&3&0.4285&4&4&4&6&3 &binary\\ \hline
3&1&13&4&1&0.0769&13&$6^*$&5&6&3 &binary\\
 & &  & &(6)&(0.4615)&(6)&( ) & & & &3-ary\\ \hline
2&2&21&5&11&0.5238&6&6&6&6&3 &binary\\ \hline
5&1&31&6&1&0.0322&31&$10^*$&7&6&3 &binary\\
 & &  & &(15)&(0.4838)&($10^*$)& ( )& & & &5-ary\\ \hline
7&1&57&8&1&0.017&57&$16^{*}$&9&6&3 &binary\\
 & &  & &(28)&(0.4912)&($14^*$)& ( )& & & &7-ary\\ \hline
2&3&73&9&45&0.6164&10&10&10&6&3 &binary\\ \hline
3&2&91&10&1&0.0109&91&$\ge 11$&11&6&3 &binary\\
 & &  &  &(54)&(0.5934)&($15^*$)&( ) & & & &3-ary\\ \hline
2&4&273&17&191&0.6996&18&18&18&6&3 &binary\\ \hline
5&2&651&26&1&0.0015&651&$\ge 27$&27&6&3 &binary\\
 & &   &  &(425)&(0.6528)&($56^*$)&( ) & & & &5-ary\\
\hline
\end{tabular}
\caption{Summary of Type II, $\ell = 3$ code parameters.}
\label{table_typeII3}
\end{center}
\vspace{-0.35in}
\begin{center}
\begin{tabular}{|c|c|c|c|c|c|c|c|c|c|c|c|}
\hline
$p$&$s$&block length&degree&dimension&rate&$d_{\min}$&$w_{\min}$&tree&girth&diameter&code\\
 &   &$n$=$1$+$p^s$+$p^{2s}$+$p^{3s}$&$d$=$p^s$+$1$& & & & &lower-bound&$g$&$\delta$&alphabet\\
\hline \hline
2&1&15&3&5&0.3333&6&6&6&8&4 &binary\\ \hline
3&1&40&4&15&0.3750&10&$\ge 8$&8&$8$&4 &binary\\
 & &  & &(15)&(0.3750)&($10^*$)& ( ) & & & &3-ary\\ \hline
2&2&85&5&35&0.4117&10&10&10&8& 4 &binary\\ \hline
5&1&156&6&65&0.4167&$20^*$&$\ge 12$&12&8& 4&binary\\
 & &   & &(65)&(0.4167)&($26^*$)& ( )& & & &5-ary\\ \hline
7&1&400&8&175&0.4375&$66^*$&$\ge 16$&16&8&4 &binary\\
 & &   & &(175)&(0.4375)&($119^*$)&( ) & & & &7-ary\\ \hline
2&3&585&9&287&0.4905&18&18&18&8&4 &binary\\ \hline
3&2&820&10&369&0.4500&$112^*$&$\ge 20$&20&8&4 &binary\\ 
 & &   &  &(395)&(0.4817)&($96^*$)&( ) & & & &3-ary\\ \hline
\end{tabular}
\caption{Summary of Type II, $\ell = 4$ code parameters.}
\label{table_typeII4}
\end{center}
\end{table}
}

{\begin{center}
\begin{figure}
 \centering{\resizebox{5.25in}{3.275in}{\includegraphics{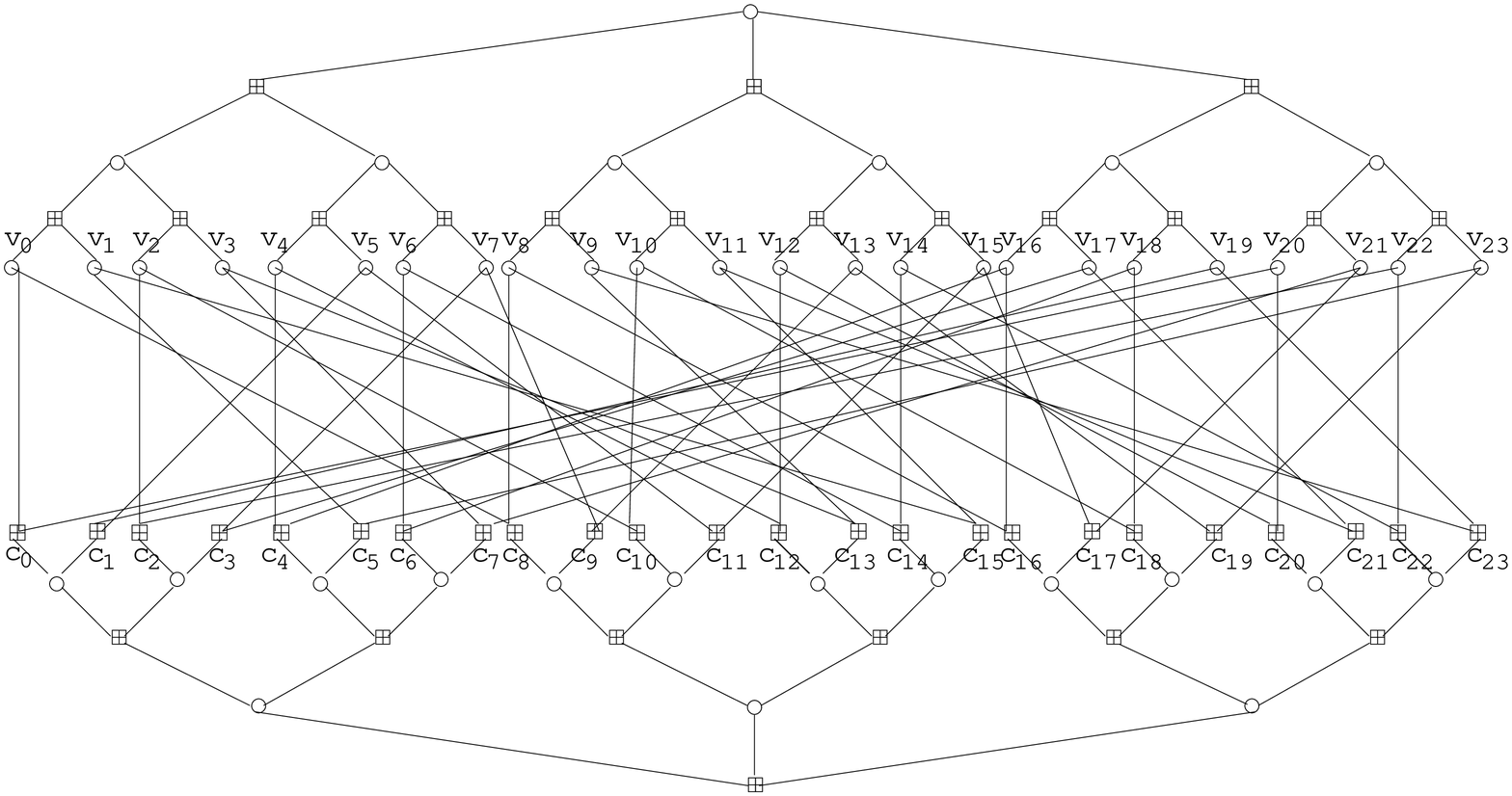}}}
\caption{Type I-A LDPC constraint graph having degree $d=3$ and girth $g=10$.}
\label{type1Ad3g10_graph}
\end{figure}
\end{center}}

\vspace{0.2in}

{\begin{center}
\begin{figure}
\centering{\resizebox{5.25in}{3.2in}{\includegraphics{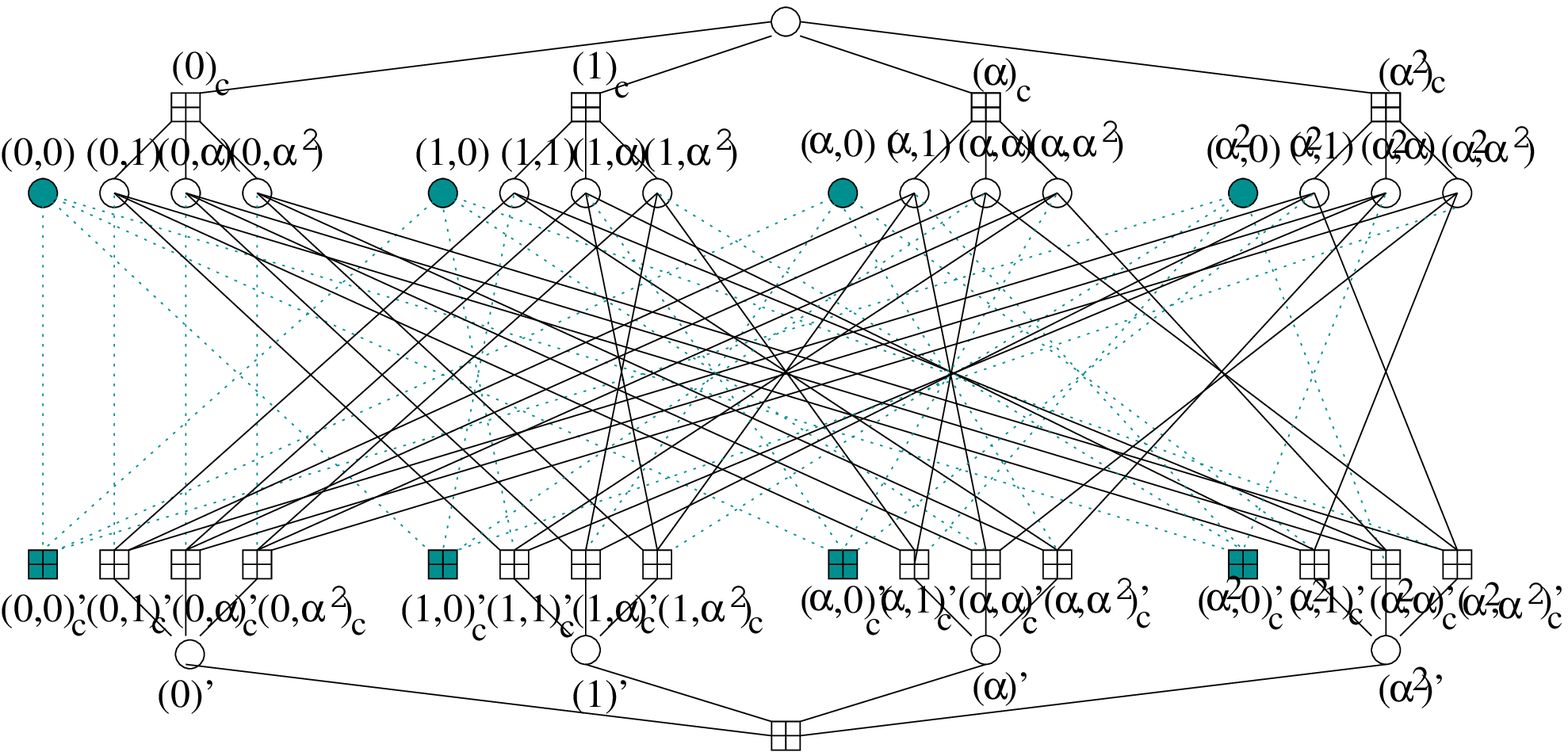}}}
\caption{Type I-B LDPC constraint graph having degree $d=4$ and girth $g=6$.}
\label{type1Bd4g6_graph}
\end{figure}
\end{center}}

\vspace{0.2in}

{\begin{center}
\begin{figure}
\centering{\resizebox{6in}{2.5in}{\includegraphics{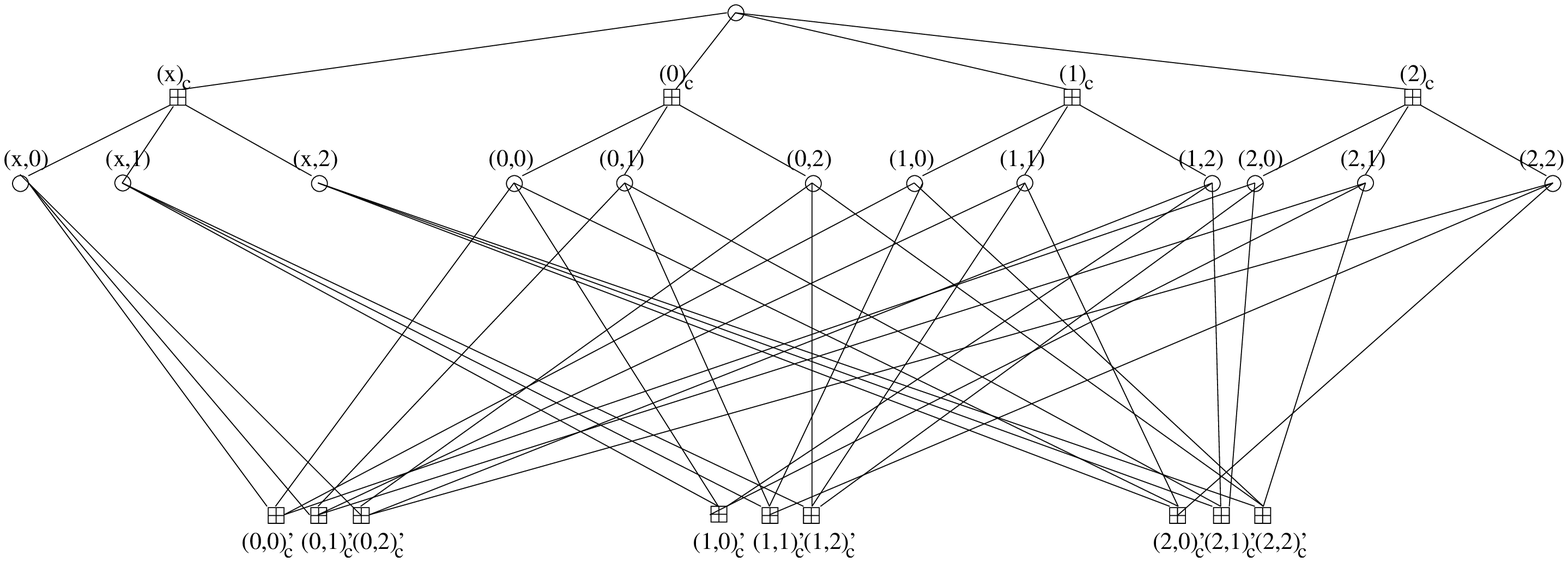}}}
\caption{Type II LDPC constraint graph having degree $d=4$ and girth $g=6$.}
\label{type2g6d4_graph}
\end{figure}
\end{center}}

\vspace{0.2in}

{\begin{center}
\begin{figure}
  \centering{\resizebox{6in}{2.5in}{\includegraphics{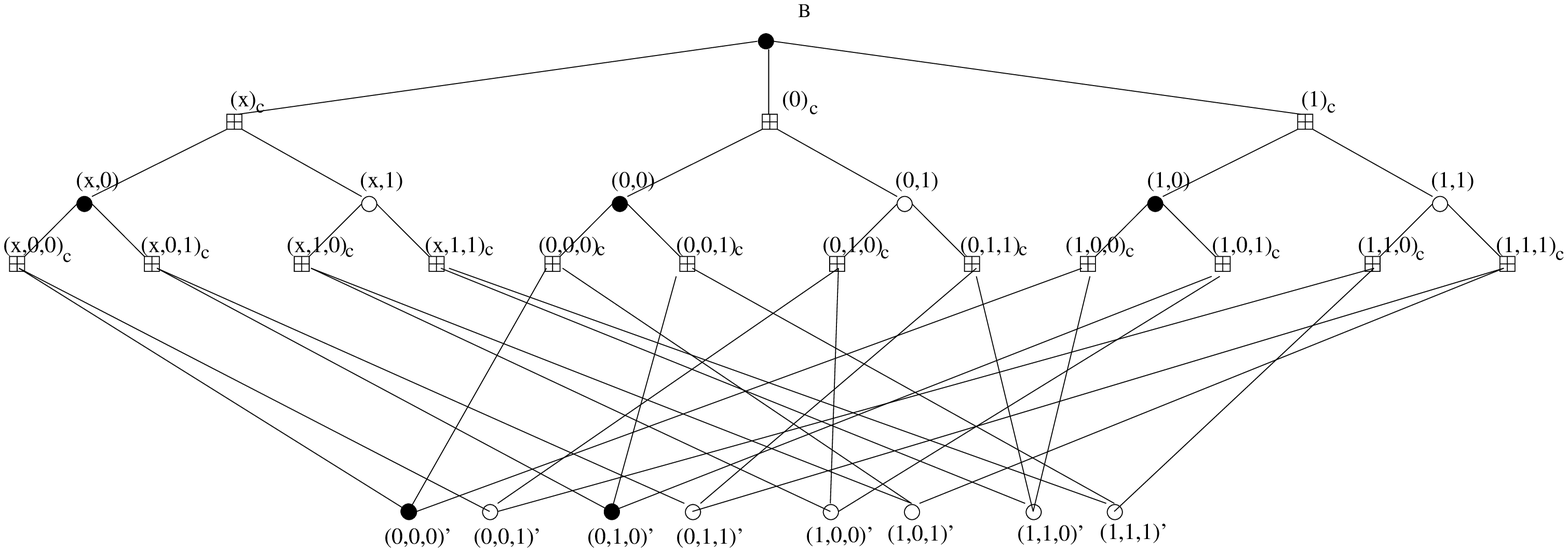}}}
\caption{Type II LDPC constraint graph having degree $d=3$ and girth $g=8$. (Shaded nodes highlight a minimum weight codeword.)}
\label{type2g8d3_graph}
\end{figure}
\end{center}}

\vspace{0.2in}

\begin{figure}
\centering{\resizebox{2.5in}{2.3in}{\includegraphics{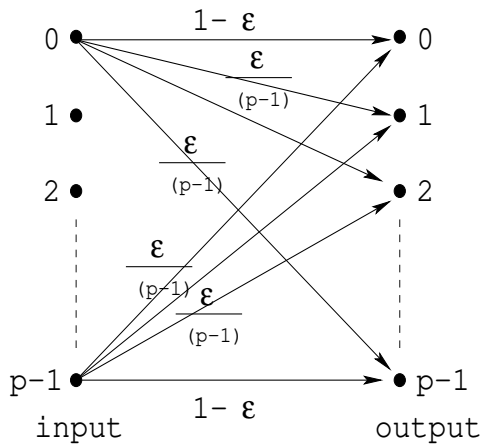}}}
\caption{A $p$-ary symmetric channel.}
\label{q_ary_sym}
\end{figure}

\vspace{0.2in}


{\begin{center}
\begin{figure}
\centering{\resizebox{4.95in}{3.5in}{\includegraphics{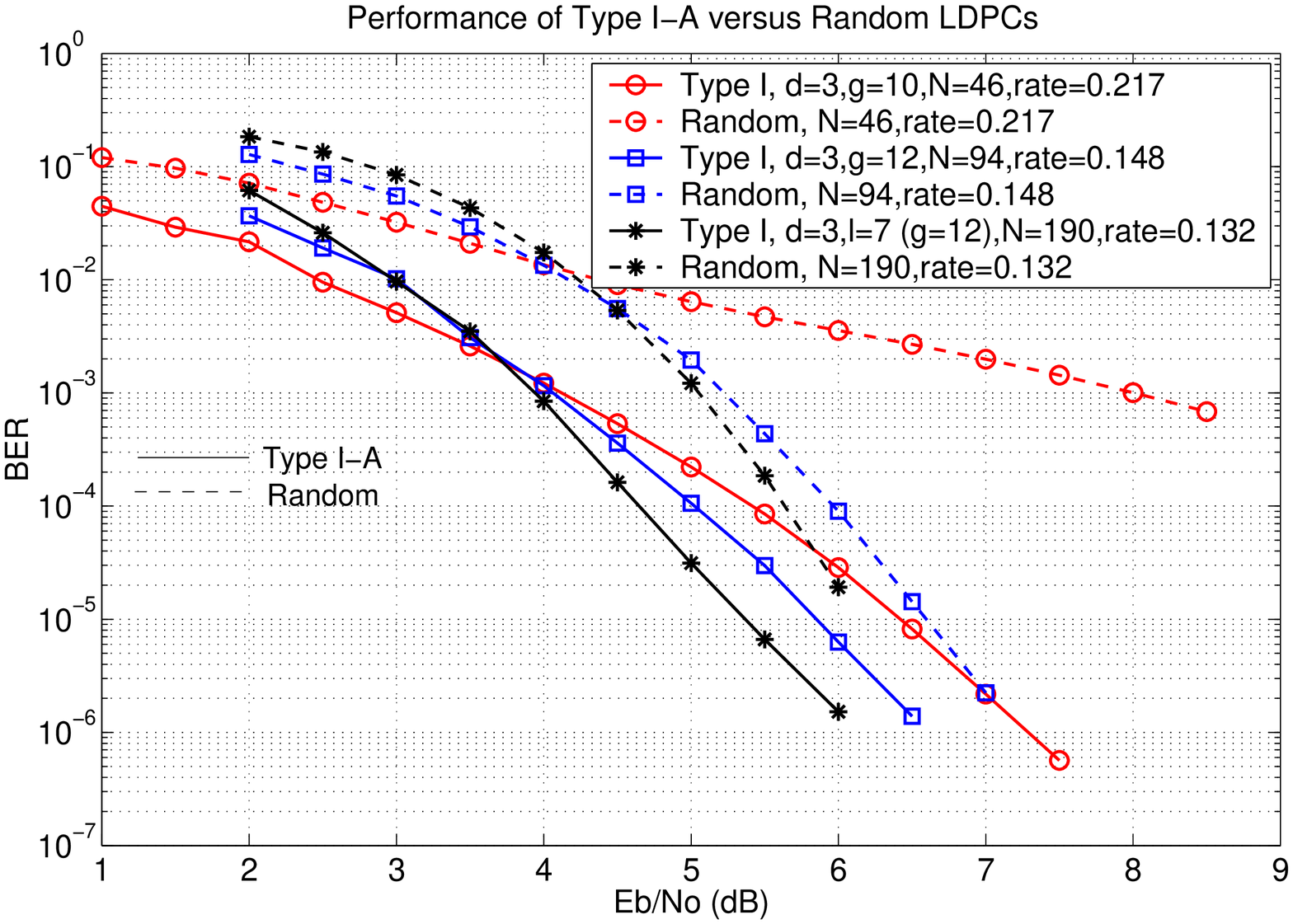}}}
\caption{Performance of Type I-A versus Random LDPC codes on the
  BIAWGNC with min-sum iterative decoding.} 
\label{type1_perf}
\end{figure}
\end{center}}

{\begin{center}
\begin{figure}
\centering{\resizebox{4.95in}{3.5in}{\includegraphics{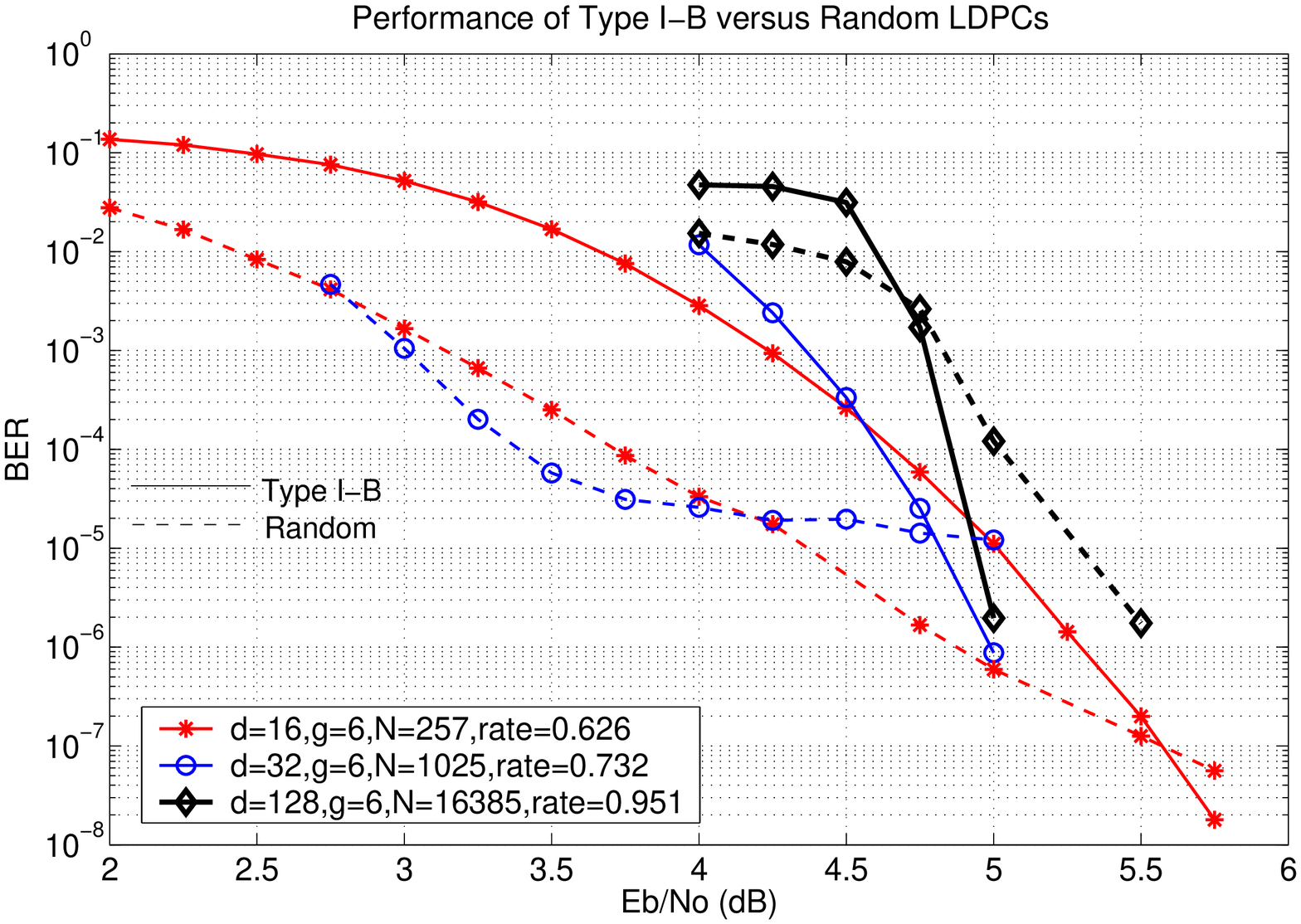}}}
\caption{Performance of Type I-B versus Random LDPC codes on the BIAWGNC with min-sum iterative decoding.}
\label{type1B_perf}
\end{figure}
\end{center}}

{\begin{center}
\begin{figure}
\centering{\resizebox{4.95in}{3.5in}{\includegraphics{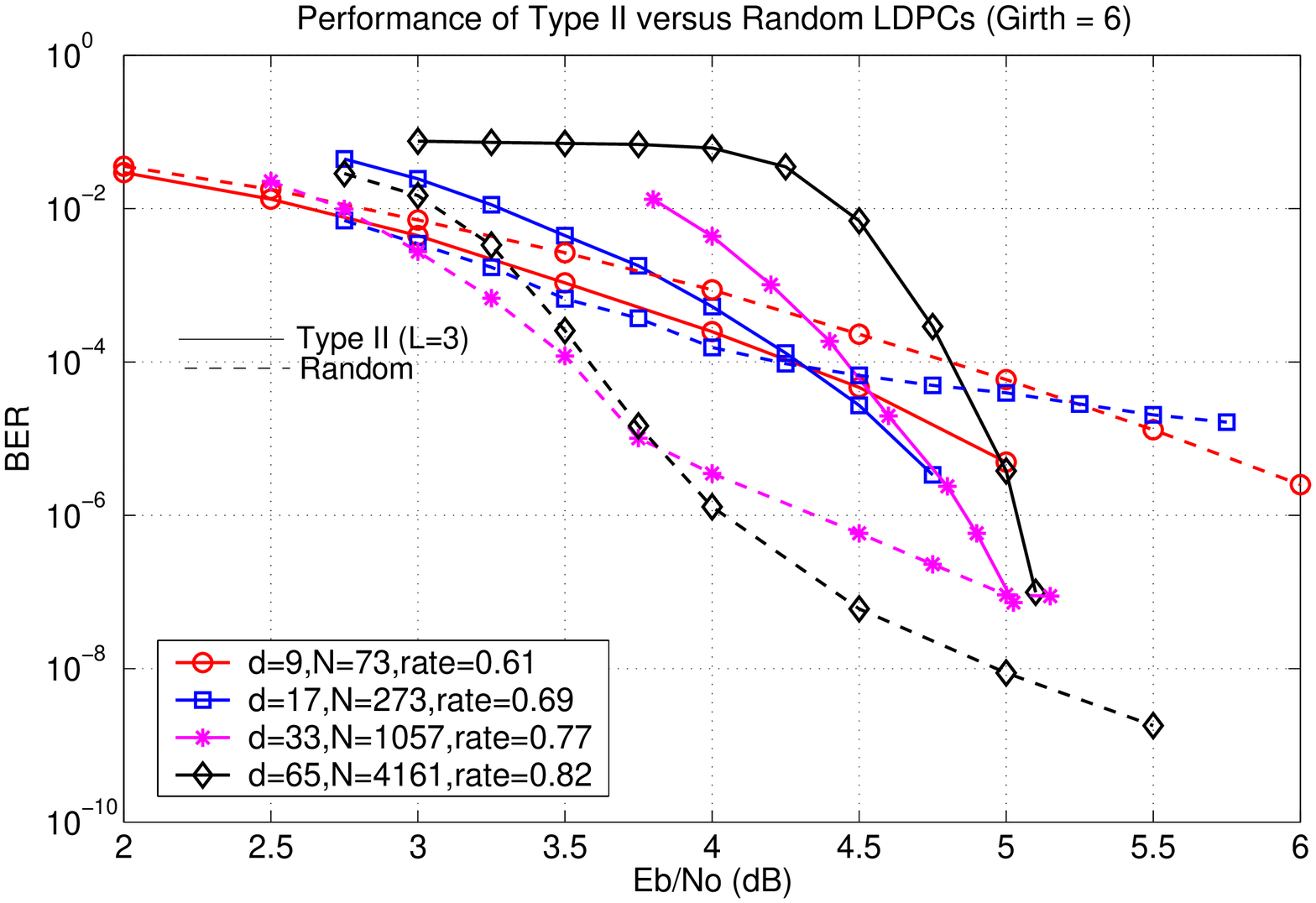}}}
\caption{Performance of Type II $\ell=3$ versus Random LDPC codes on the BIAWGNC with min-sum iterative decoding.}
\label{type2g6_perf}
\end{figure}
\end{center}}
{\begin{center}
\begin{figure}
\centering{\resizebox{4.95in}{3.5in}{\includegraphics{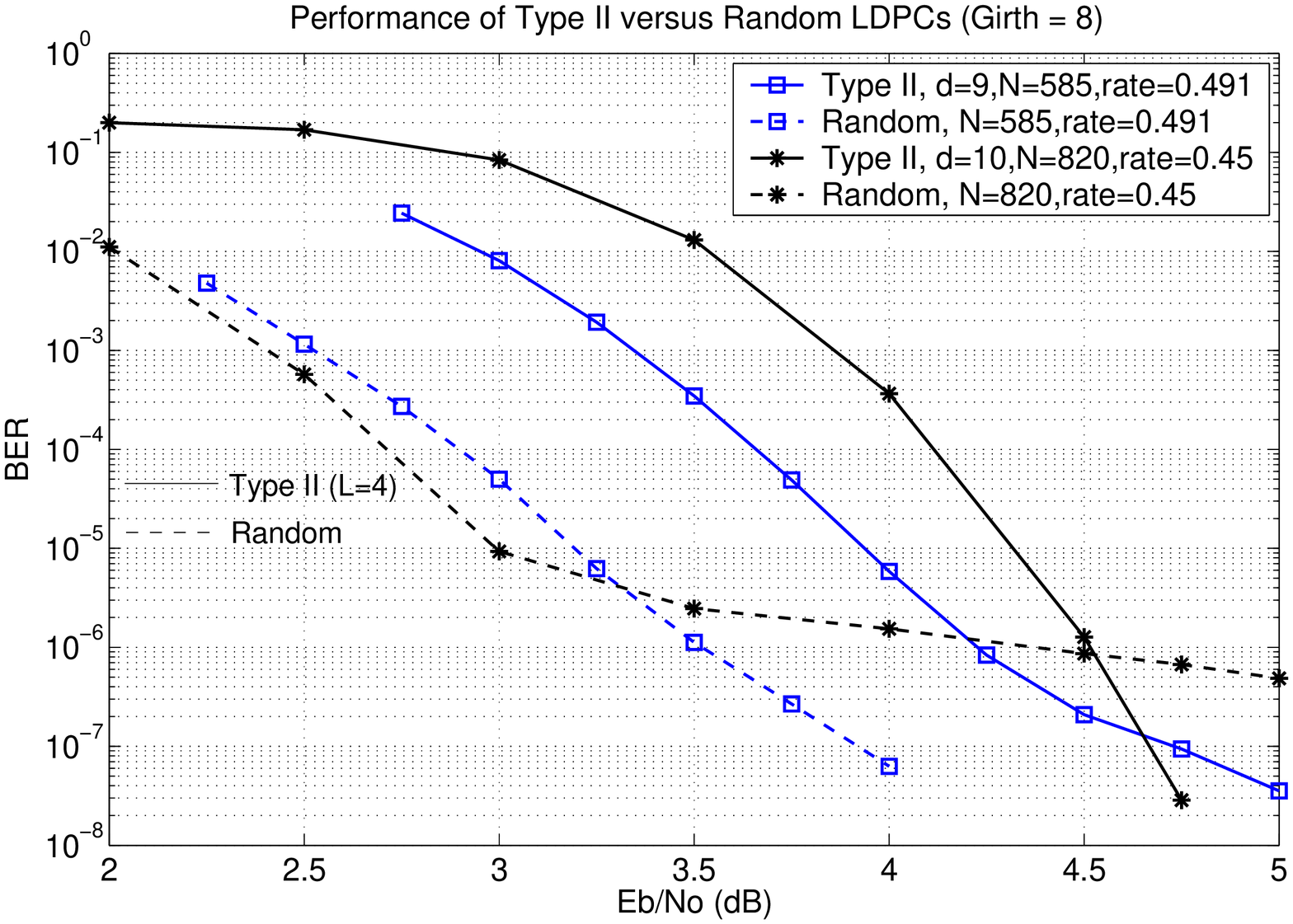}}}
\caption{Performance of Type II $\ell=4$ versus Random LDPC codes on the BIAWGNC with min-sum iterative decoding.}
\label{type2g8_perf}
\end{figure}
\end{center}}

{\begin{center}
\begin{figure}
\centering{\resizebox{4.95in}{3.5in}{\includegraphics{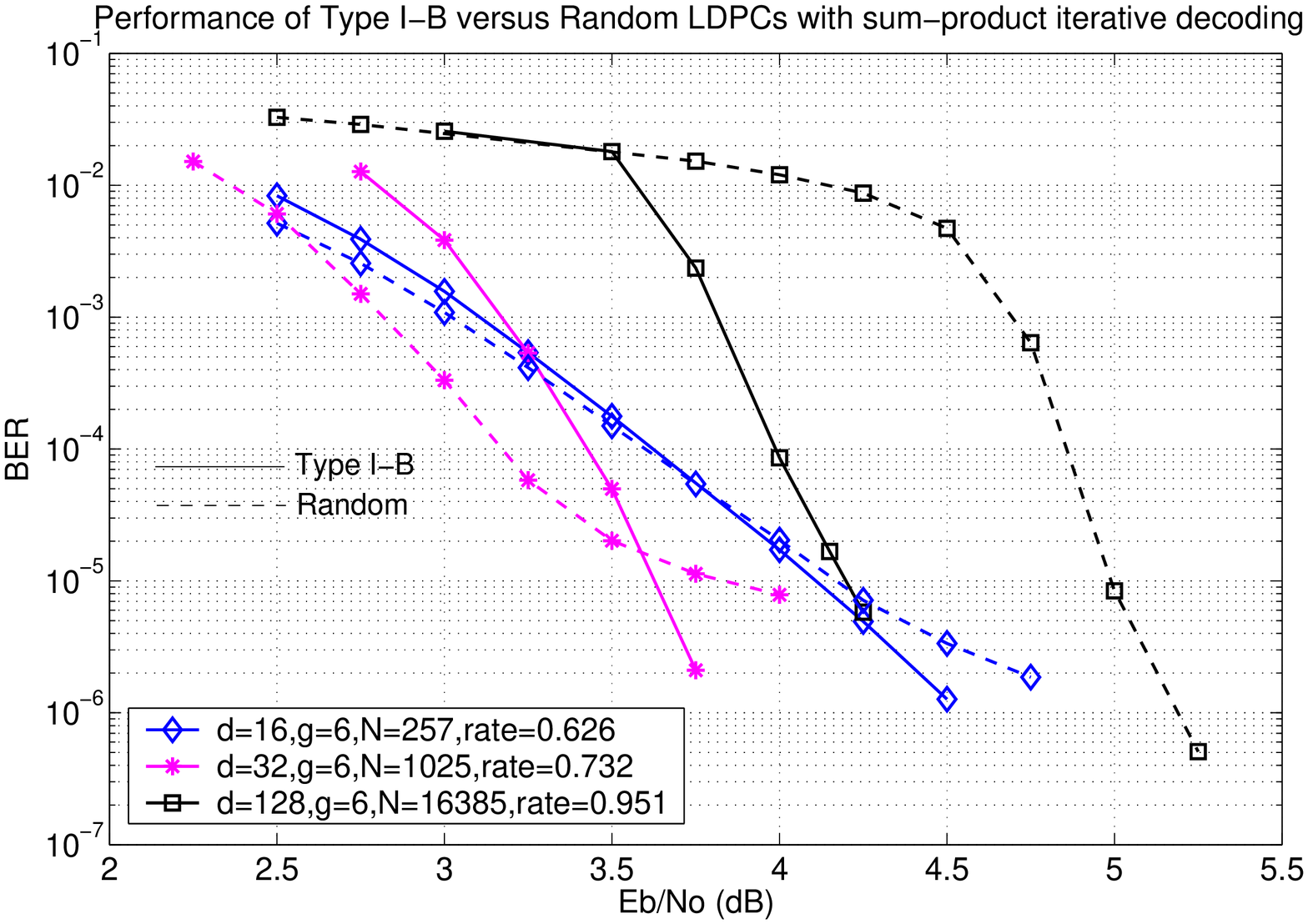}}}
\caption{Performance of Type I-B versus Random LDPC codes on the BIAWGNC with sum-product iterative decoding.}
\label{type1B_SPperf}
\end{figure}
\end{center}}

{\begin{center}
\begin{figure}
\centering{\resizebox{4.95in}{3.5in}{\includegraphics{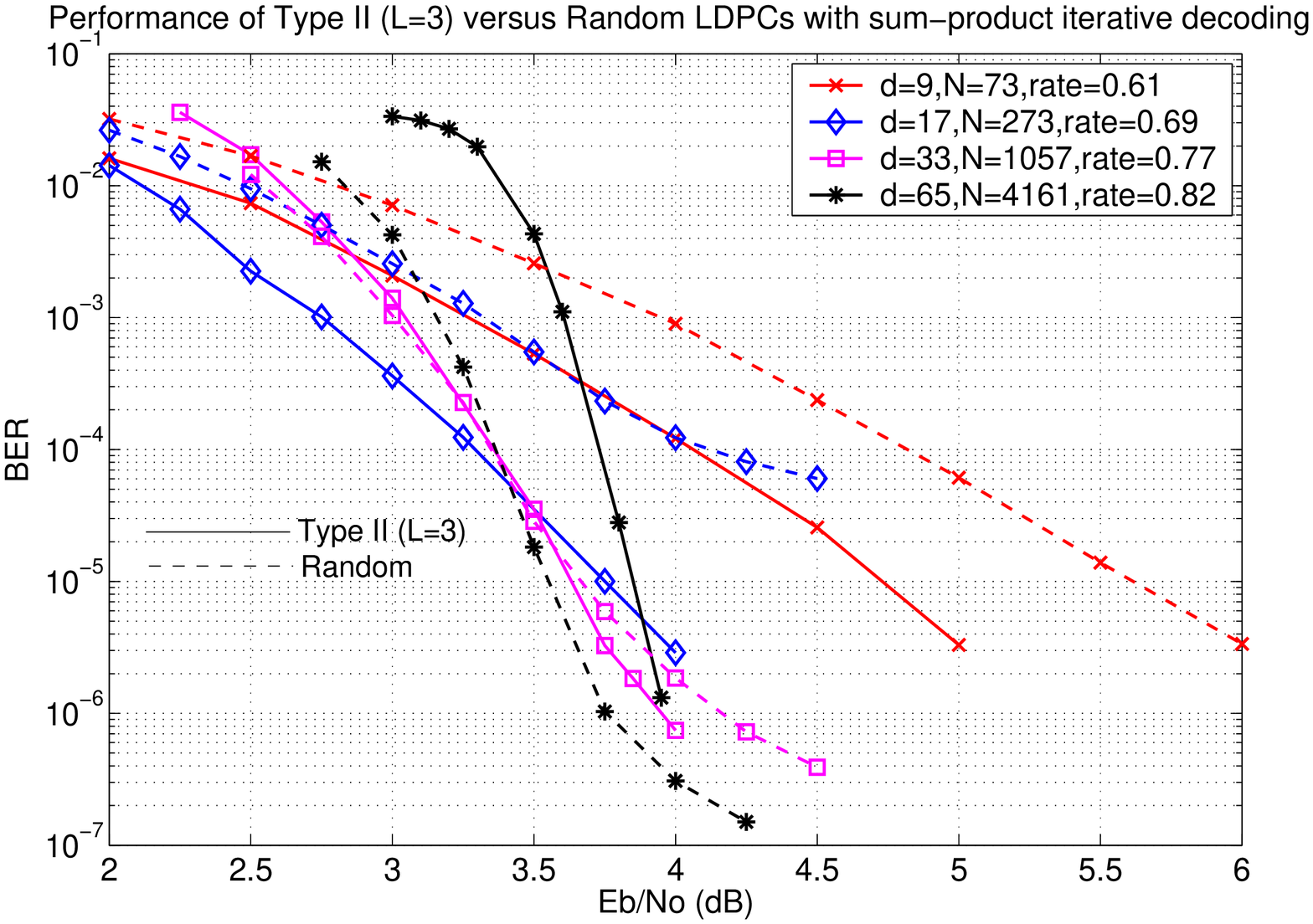}}}
\caption{Performance of Type II $\ell=3$ versus Random LDPC codes on the BIAWGNC with sum-product iterative decoding.}
\label{type2l3_SPperf}
\end{figure}
\end{center}}
{\begin{center}
\begin{figure}
\centering{\resizebox{4.95in}{3.5in}{\includegraphics{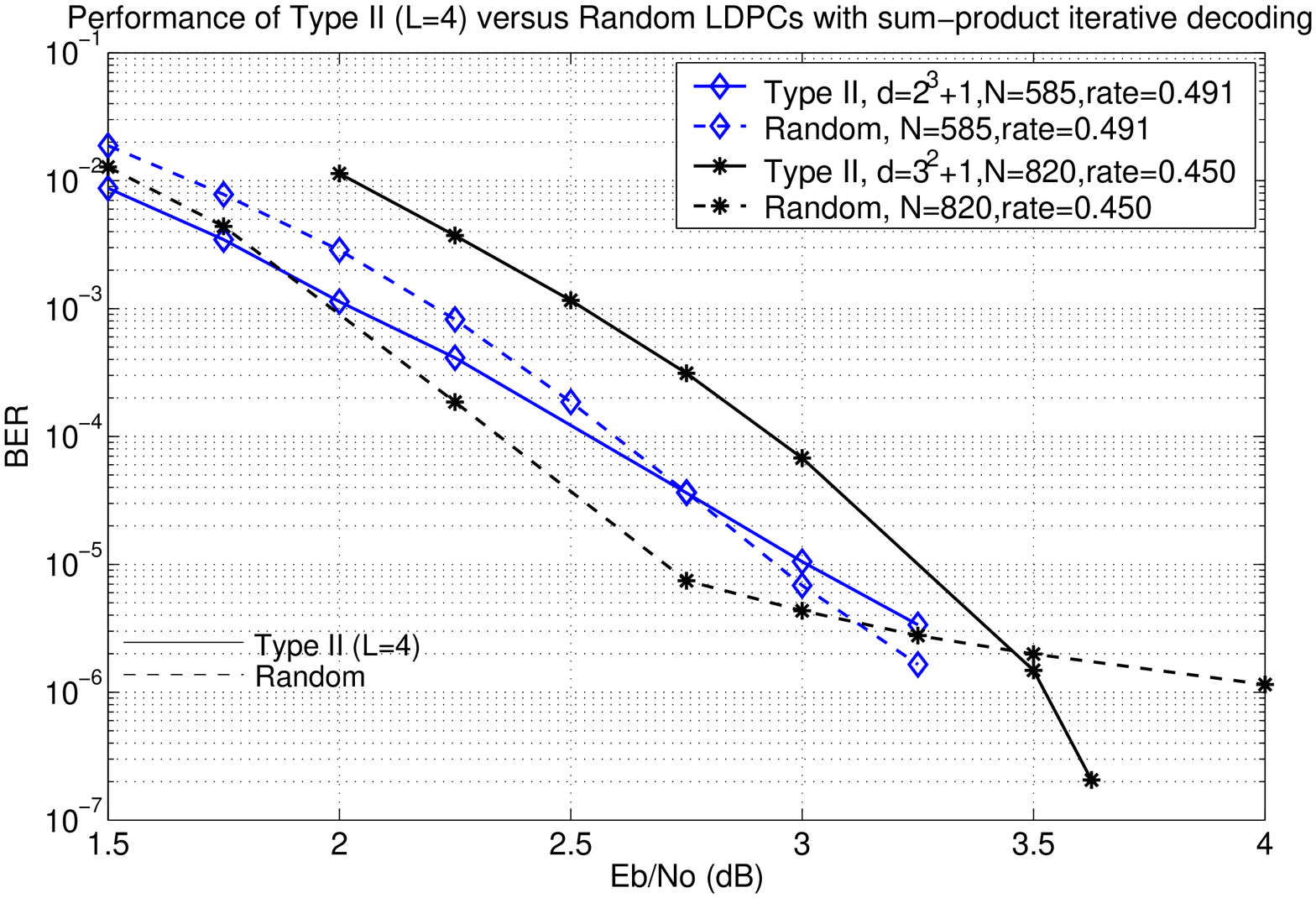}}}
\caption{Performance of Type II $\ell=4$ versus Random LDPC codes on the BIAWGNC with sum-product iterative decoding.}
\label{type2l4_SPperf}
\end{figure}
\end{center}}

{\begin{center}
\begin{figure}
\centering{\resizebox{4.95in}{3.5in}{\includegraphics{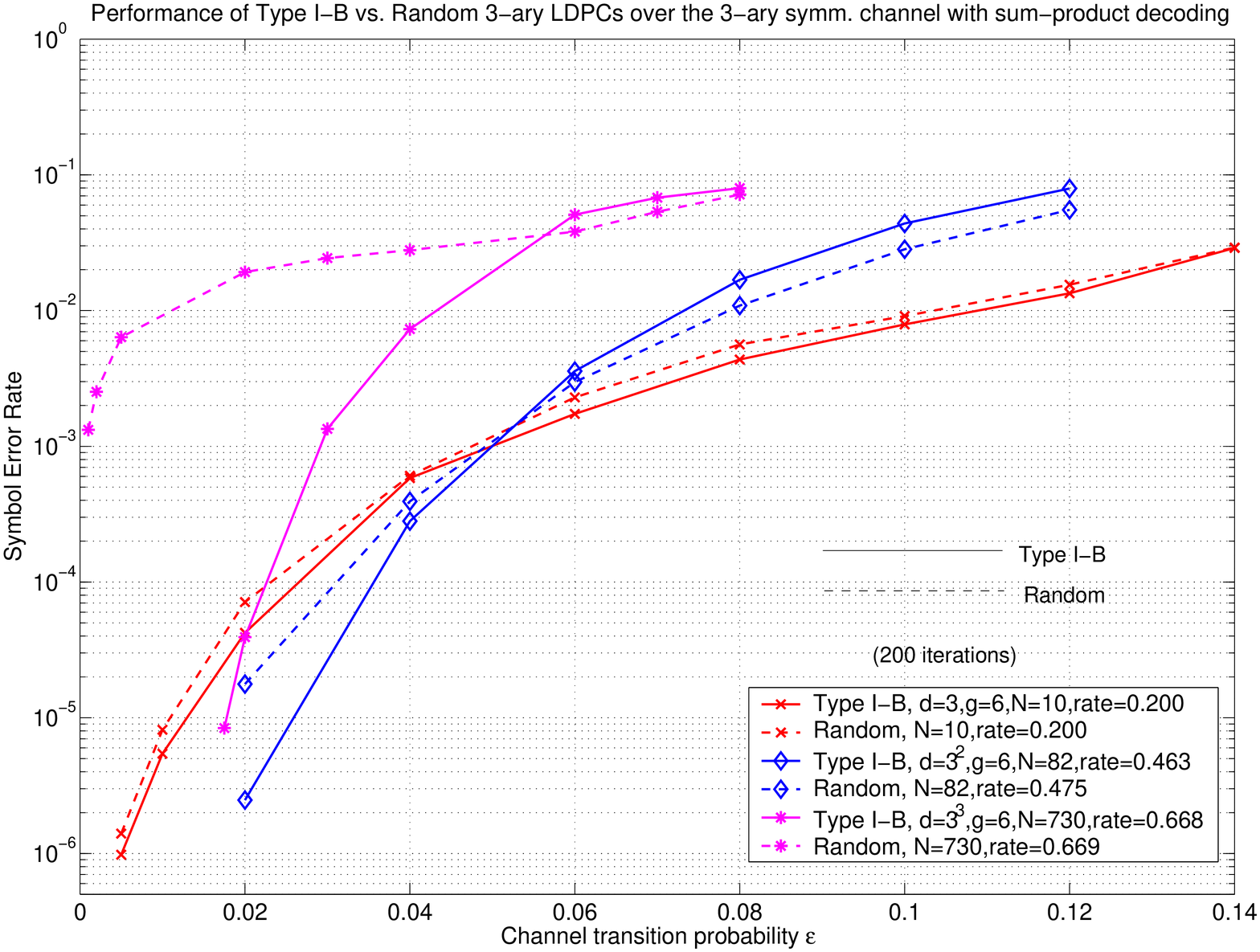}}}
\caption{Performance of Type I-B versus Random 3-ary LDPC codes
  on the 3-ary symmetric channel with  sum-product iterative decoding.}
\label{type1B_pary_perf}
\end{figure}
%
\begin{figure}
\centering{\resizebox{4.95in}{3.5in}{\includegraphics{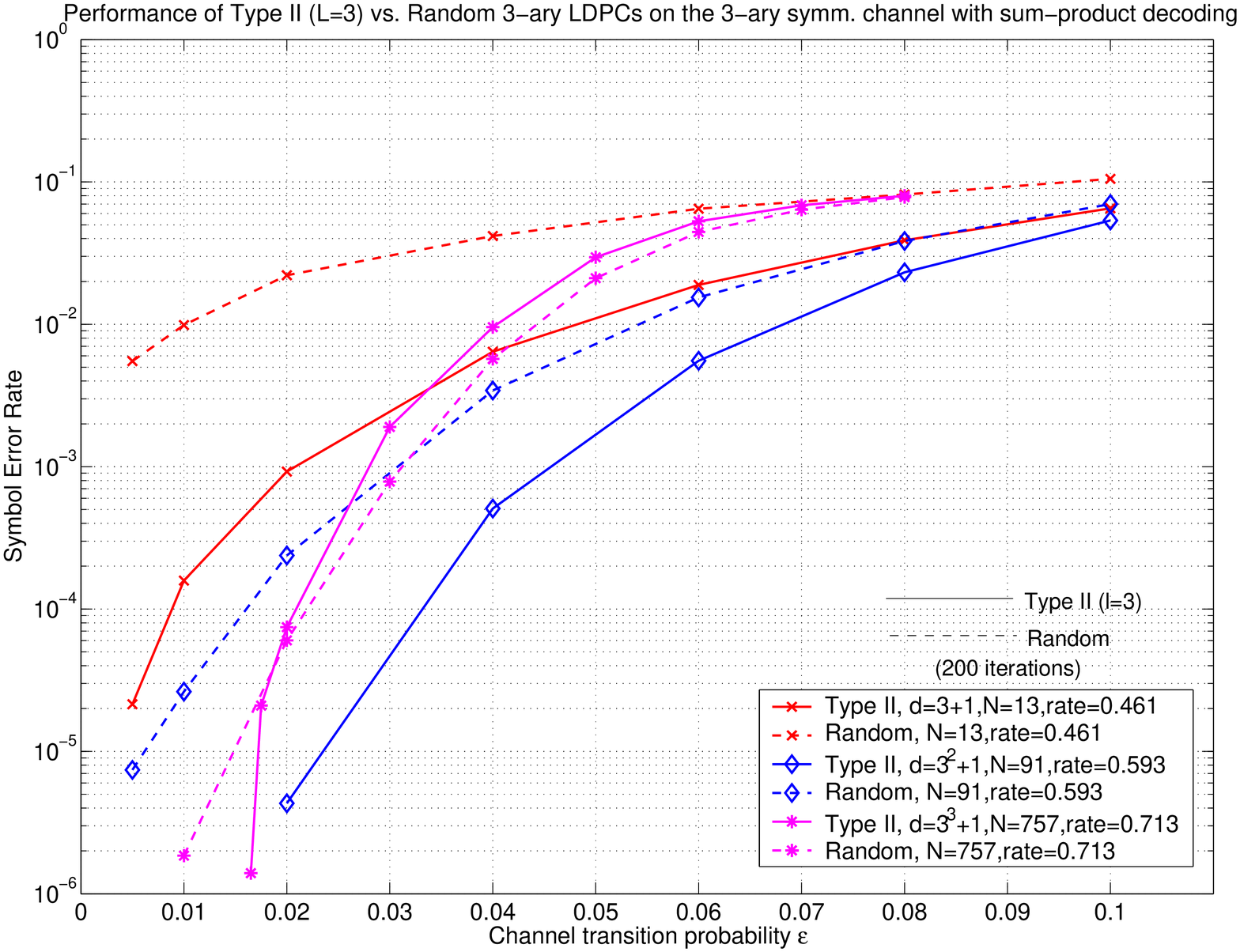}}}
\caption{Performance of Type II $\ell=3$ versus Random 3-ary LDPC codes on the 3-ary symmetric channel with sum-product iterative decoding.}
\label{type2l3_pary_perf}
\end{figure}
\end{center}}

{\begin{center}
\begin{figure}
\centering{\resizebox{4.95in}{3.5in}{\includegraphics{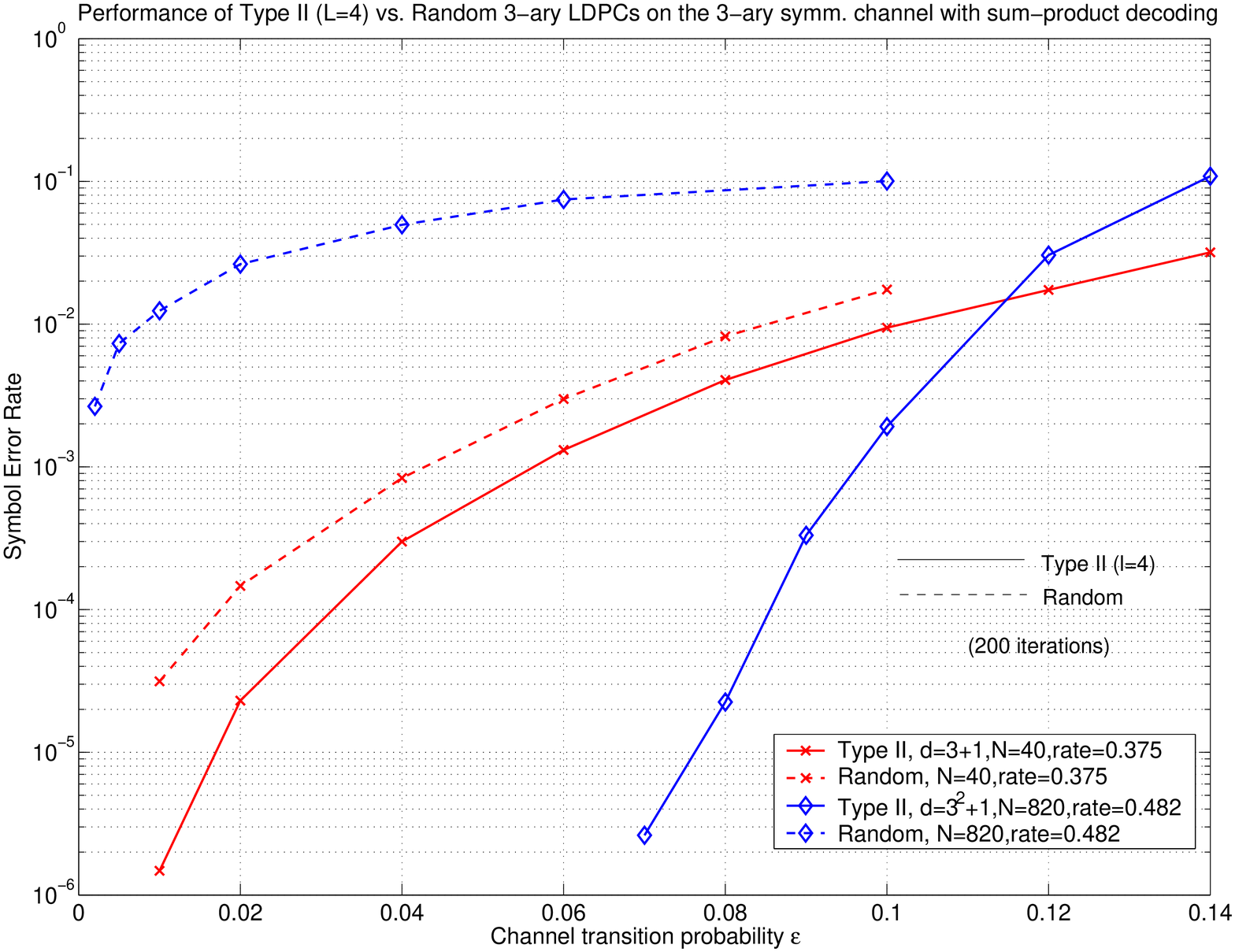}}}
\caption{Performance of Type II $\ell=4$ versus Random 3-ary LDPC codes on the 3-ary symmetric channel with sum-product iterative decoding.}
\label{type2l4_pary_perf}
\end{figure}
\end{center}}

\end{document}